\documentclass[]{spie}  
\usepackage[left=1in,top=1in,right=1in,bottom=1in,footskip = 0.333in]{geometry}
 
\usepackage{mathtools} 
\usepackage{amsmath,amsfonts,amssymb}
\usepackage{graphicx}
\graphicspath{{Media}}
\usepackage[colorlinks=true, allcolors=blue]{hyperref}

\usepackage{tabularx} 
\newcolumntype{b}{X}
\newcolumntype{s}{>{\hsize=.4\hsize}X}
\newcolumntype{L}[1]{>{\raggedright\let\newline\\\arraybackslash\hspace{0pt}}m{#1}} 
\newcommand{\magasec}{mag\,asec$^{-2}$}

 \newcommand{\suchthat}{\mid}

\newcommand{\std}{\operatorname{Std}}
\newcommand{\var}{\operatorname{Var}}

\newcommand{\Deff}{D_\textrm{eff}}
\newcommand{\Aeff}{A_\textrm{eff}}
\newcommand{\tputWord}{\textrm{Throughput}}
\newcommand{\tput}{\tau}
\newcommand{\diam}{D}
\newcommand{\contrast}{C}
\newcommand{\iwa}{\iota}
\newcommand{\mode}{\textrm{mode}}
\newcommand{\grid}{\textrm{grid}}
\newcommand{\sobol}{\textrm{sobol}}
\newcommand{\pool}{\textrm{pool}} 

\title{HWO Yield Sensitivities in the NIR and NUV}

\author[a]{Rhonda Morgan}
\author[b,c]{Dmitry Savransky}
\author[a]{Michael Turmon}
\author[a]{Mario Damiano}
\author[a]{Renyu Hu}
\author[a]{Bertrand Mennesson}
\author[a]{Eric E. Mamajek}
\author[d]{Tyler D. Robinson}
\author[a]{Armen Tokadjian}

\affil[a]{Jet Propulsion Laboratory, California Institute of Technology, 4800 Oak Grove Dr., Pasadena, CA, 91109}
\affil[b]{Sibley School of Mechanical and Aerospace Engineering, Cornell University, Ithaca, NY, 14853}
\affil[c]{Carl Sagan Institute, Cornell University, Ithaca, NY, 14853}
\affil[d]{Lunar and Planetary Laboratory, University of Arizona, Tucson, AZ, 85721}

\authorinfo{Send correspondence to R.M.: rhonda.morgan@jpl.nasa.gov}

\begin{document} 
\maketitle

\begin{abstract}

Habitable Worlds Observatory (HWO) will search for biosignatures from Earth-size exoplanets in the habitable zones of nearby stars. The wavelength range for biosignatures used by the HabEx and LUVOIR mission concept studies was 200 nm to 2 microns and, as such, this is a candidate wavelength range for HWO.  The visible wavelength range (500-1000 nm) provides for detection of water, oxygen, and Raleigh scattering; the near-ultraviolet is valuable for detection of ozone; and the near-infrared enables detection of carbon dioxide and methane for Earth-like atmospheres.  Damiano et al. 2023 showed the significant improvement in spectral retrieval reliability when the NUV and NIR are both used with the visible. However, the challenge of the NUV, in addition to the technological and engineering challenges of starlight suppression in the NUV, is the drop in flux of host stars. In the NIR, the challenge is the geometric access to the habitable zone due to the wavelength dependency of the inner working angle limit of coronagraphs. For these reasons, exoplanet yields are lower in the NUV and NIR than in the visible \cite{morgan2023exo,morgan2024exo} and some instrument parameters are more critical for improving NUV and NIR yields than others. In this paper we present a new capability for performing a large number of end-to-end yield modeling simulations to enable large, multivariate parameter sweeps.  We utilize this capability to calculate the Visible, NIR, and NUV yield sensitivities to the instrument parameters: aperture diameter,  coronagraph core throughput, contrast, and inner working angle (IWA). We find that parameter interactions are important in determining yield, the most important of which is the interaction between contrast and IWA, but that the  strength of that interaction is different in each of the three wavebands. 

\end{abstract}

\section{INTRODUCTION}
\label{sec:intro}  

The Astro2020 Decadal Survey recommended a ``future large IR/O/UV telescope optimized for observing habitable exoplanets and general astrophysics'' to be mature in concept by end of the decade and to be launched in the early 2040's. Astro2020 put forward a mission ``capable of surveying a hundred or more nearby Sun-like stars to discover their planetary systems and determine their orbits and basic properties'' and recommended as the exoplanet science goal ``to search for biosignatures from a robust number of $\sim$25 habitable zone [exo]planets.'' The number of $\sim$25 habitable zone (HZ) exoplanets for a $\sim$6 m telescope arises from Astro2020 Figure 7.6, in which yield is the number of HZ Earth-sized exoplanets spectrally characterized to search for water, specifically the deep water line at 940 nm, using a 20\% bandpass spectra from 800-1000 nm in wavelength with spectral resolution of 70 and signal-to-noise ratio (SNR) of 5\cite{stark2015lower,luvoir2019, morgan2019sdetfinal}.  This spectra represents the shortest integration time investment for a minimum-quality spectra to determine if the HZ exoplanet has an interesting atmosphere that could be worth following up with additional spectral observations in other 20\% sub-bands. The determination of yield in other sub-bands for biosignatures will require higher quality spectra, and thus longer integration times and longer mission times. It will also require spectra in the NIR and NUV.

Recent work by Damiano et al. \cite{damiano2021reflected, damiano2022reflected, damiano2023reflected} showed that the visible, NUV, and NIR wavelengths are necessary to understand the spectral retrievals of small rocky planet atmospheres.  Damiano et al. 2022 showed that the visible spectrum was sufficient to an Archean Earth atmosphere, but that without the NIR ($1-1.8 \mu m$), the Modern Earth atmosphere, a Venus-like atmosphere, and Mars-like atmosphere will appear to be correctly modeled in the visible but will be highly inaccurate in the NIR.  Damiano et al. 2023 showed that the NUV (250 - 350 nm) provided proper constraints on $O_3$ for Proterozoic Earth atmosphere with 1\% and 0.1\% of the $O_2$ of Modern Earth, even without the NIR (though the NIR spectrum was incorrect) and that $SO_2$ does not cause confusion of $O_3$ detection. For Proterozoic Earth with visible + NIR and no NUV spectra, the $O_3$ could be detected, but with significantly reduced sensitivity and could not set constraints on abundances. Damiano et al. 2023 additionally showed that the Visible + NUV do not constrain the clouds and surface parameters but that Visible + NUV + NIR can constrain clouds and surface parameters.

Since NUV and NIR are so important for understanding biosignatures and terrestrial atmospheres, it is essential to calculate yields in the NUV and NIR and to understand the sensitivity of key instrument parameters in the NUV and NIR, because they may differ from sensitivities in the visible. Towards this end, we developed a new capability to perform a large number of parameter sweeps using the end-to-end yield modeling code EXOSIMS. We used this capability to perform an initial study of sensitivities in the visible, NIR, and NUV to four key parameters: aperture diameter, coronagraph contrast, IWA, and core throughput. We included all of the cross terms of the input parameters, i.e. the study was multivariate, resulting in 960 instrument configurations, where each instrument configuration was evaluated with a Monte Carlo ensemble of 100 mission simulations. Then we performed two types of analysis on the yield results to create models of sensitivity.  The first analysis was a statistical fit to a parametric model. The second analysis was a random forest analysis to develop a machine learning model.



Section~\ref{sec:approach} describes astrophysical parameters,
our approach for computing yields, 
and the design of this parametric study. It is important to consider all yields as relative, not absolute, due to the large uncertainties in the occurrence rate of exo-Earths.
Section~\ref{sec:results} describes the properties of the estimated
yields we obtained.
The sensitivity of these yields to parameter variations are described in 
Section~\ref{sec:analysis},
and we offer conclusions in Section~\ref{sec:conclusion}.


\begin{table}
\caption{Spectral characterization metrics
\label{tbl:passbands}}
\centering
\begin{tabularx}{.46\textwidth}{ccccc}
    \hline\hline
    Label & Species & Wavelength\textsuperscript{a} & R\textsubscript{S} & SNR\\
    \hline
    Visible & H\textsubscript{2}0 & 800-1000 nm & 70 & 5\\
    NIR & CO\textsubscript{2} & 1350-1650 nm & 40 & 8.5\\
    NUV & O\textsubscript{3} & 250 -310 nm & 7 & 5 \\
    \hline
    \multicolumn{5}{l}{\textsuperscript{a}\footnotesize{Wavelength range is a single 20\% passband}}\\
\end{tabularx}
\end{table}

\section{Approach}
\label{sec:approach}

We performed end-to-end yield simulations for selected parameter combinations, following 
two separate experimental designs, to study the influence of key parameters and parameter interactions. 
For each parameter combination in each experimental design, we used EXOSIMS to generate
a Monte Carlo ensemble of $N_e = 100$ mission simulations, thereby generating an
expected yield, and an estimate of its Monte Carlo error, for that parameter tuple.

\subsection{Experimental Design}
Our main experiment used a multivariate parameter sweep in
a full-factorial design across the following grid points:
\begin{itemize}
\itemsep 0pt 
\item Diameter $\diam \in \{ 6, 7, 8, 9 \}$ m
\item Throughput $\tput \in \{ 0.1, 0.2, 0.4, 0.6 \}$\pagebreak[4]
\item IWA $\iwa \in \{ 20, 37, 54, 70 \}$ mas
\item Contrast $\contrast \in \{ 10^{-9},\, 3\cdot 10^{-10},\, 10^{-10},\, 3\cdot 10^{-11},\, 10^{-11} \}$
\item Observing bands: $ \mode \in \{ \textrm{Vis}, \textrm{NIR}, \textrm{NUV} \}$ (see Table~\ref{tbl:passbands})
\end{itemize}

An instrument configuration can be captured by a 5-tuple of parameters
\begin{equation}
    \theta = (\mode, \diam, \contrast, \tput, \iwa)
\end{equation}
and the factorial design is just the Cartesian product of the above 
single-parameter sets:
\begin{equation}
    \mathcal E_\grid = \{ \theta \} 
    = \{ D \} \times \{ \tput \} \times \{ \iwa \} \times \{ \contrast \} \times \{ \mode \}
    \quad.
\end{equation}
The total number of instrument configuration in the design 
is the product of sampling cardinalities: 
$N_s = 4 \cdot 4 \cdot 4 \cdot 5 \cdot 3 = 960$.
The full-factorial experimental design allows resolution of some 
interactions between any input parameters.

The above experimental design gives a sense of the overall yield space, 
but is restricted in only sampling the key parameters at a small number of 
tie points on the above grid.
Because one of our objectives is to understand finer-scale
parameter interactions, 
it is important to have an experimental design that samples outside the grid-constrained set.
To address this potential limitation, we generated a supplemental experimental design $\mathcal E_\sobol$ 
based on the well-known Sobol sequence\cite[sec.~2.7]{DKS2013qmc}. 
This generates a network of points, distributed in the unit hypercube, 
that is one example of a Quasi Monte Carlo (QMC) design.
It has the property that it more uniformly covers the four-dimensional parameter space than, 
for example, a pseudorandom set of points would, 
and in particular it samples parameter values outside 
of the tie points in the gridded design $\mathcal E_\grid$.
Our supplemental experimental design used the Sobol 
sequence containing $256$ distinct points in the four-dimensional
cube $[0,1]^4$.
These points were then scaled to match the endpoints of the original experimental design (with log-domain scaling for the contrast parameter).
We fixed this design, and replicated it for each of the three observing bands,
for a total number of scenarios $N_s = 256 \cdot 3 = 768$ 
in the set $\mathcal E_\sobol$.

\begin{figure}
   \begin{center}
   \begin{tabular}{c} 
   \includegraphics[width=.93\textwidth]{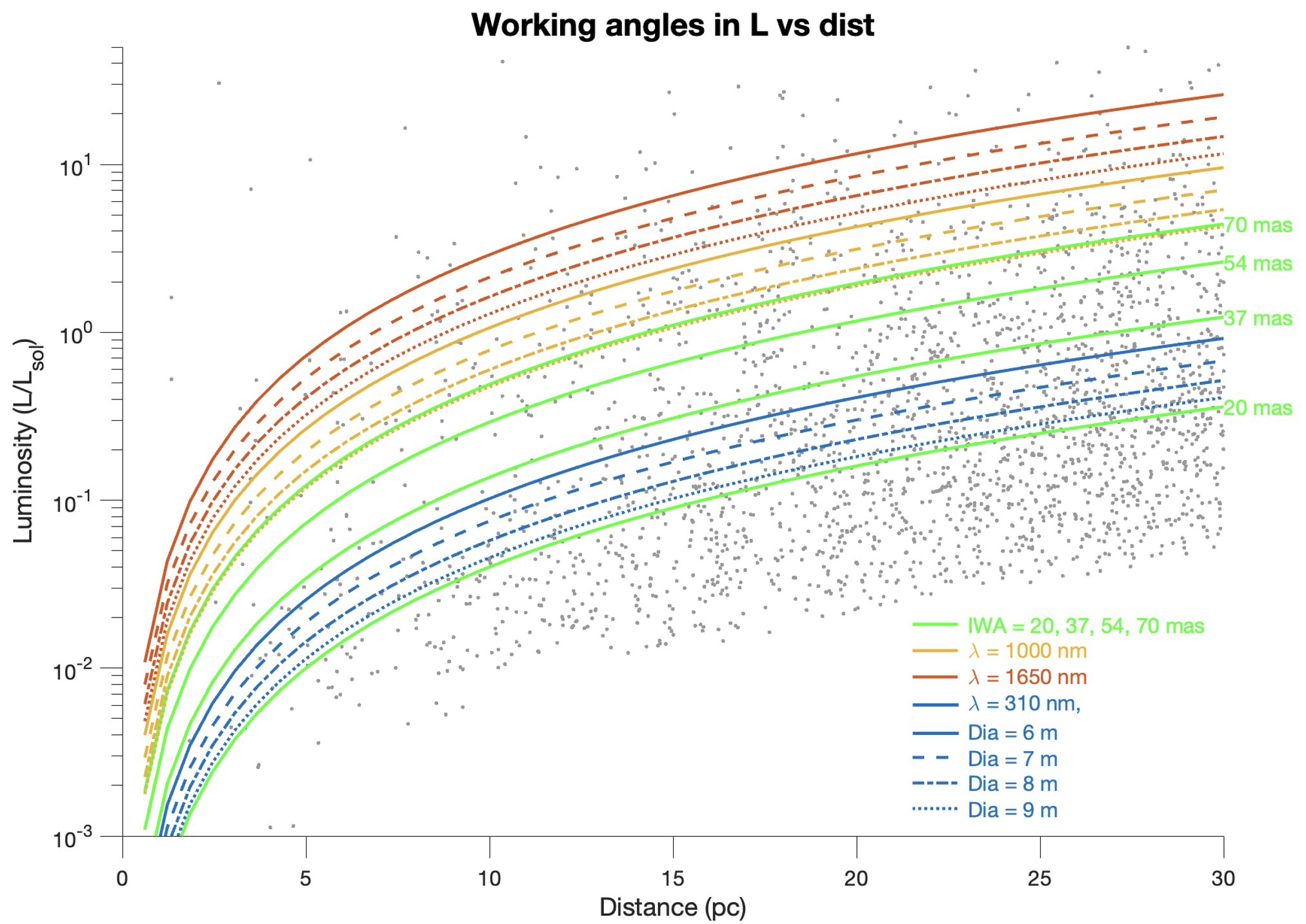}
   \end{tabular}
   \end{center}
    \caption{\label{fig:LvDist}
IWAs shown in stellar luminosity vs. distance for the input values (green) and at $3 \lambda/D$ for the input diameter and red-edge of the wavelength passbands. The visible $3 \lambda/D$ values are plotted in yellow, the NIR in red, and the NUV in blue. The aperture inscribed diameter is indicated by the line dashes: 6m is a solid line, 7 m is a dashed line, 8 m is a dash-dot line, and 9 m is a dotted line.}
\end{figure} 
The IWA parameter $\iwa$ is sampled in units of angle (milliarcsecands), 
so it is defined with respect to, and samples, 
the astrophysical space of habitable zones. IWA is more commonly defined in terms of $\lambda / D$, where $\lambda$ is the observation wavelength, as this is convenient in describing coronagraph performance which scales with $\lambda / D$. However, good sensitivity experiment design maximizes independence between input parameters, which would not happen when sampling IWA in terms of $\lambda / D$. Instead, sampling IWA in angle-space preserves a consistent population of targets across instrument configurations. Note that at 575 nm (the center of the photometric detection passband), the IWA range corresponds to a lower bound of 20 mas for a 9-m aperture at IWA = 1.5 $\lambda / D$ and an upper bound of 70 mas for a 6-m aperture with IWA = 3.5 $\lambda / D$. 

This is shown in Figure \ref{fig:LvDist}.  The green curves show the IWA samples as translated into stellar distance luminosity via Earth Equivalent Insolation Distance (EEID).  Also shown are the IWA's for a 3.5 $\lambda / D$ for each of the three wavelength passbands (yellow, blue, and red for visible, NUV, and NIR respectively) and each of the four aperture diameters by line style. Here we see the impact of the wavelength on the accessible targets. the accessible targets are above the IWA line. For the NIR (red lines), there are few targets available than at the visible (yellow lines), and the NUV (blue lines) has significantly more targets accessible. For a NIR coronagraph to reach the same targets as a visible coronagraph, the IWA must be reduced (by a different coronagraph design or a larger diameter aperture). For the NUV, the IWA can be degraded, because the targets that the NUV can access at 3.5 $\lambda / D$ are not accessible by the visible wavelength coronagraph performing the blind-search survey. An alternative observing scenario in which the blind-search is performed in the NUV could be investigated, but this would be a matter for future work.

\textbf{Assumptions:} We use the simplifying assumption that throughput and contrast are constant between the IWA and OWA. This removes the variation in coronagraph designs and focuses on major trends. We use OWA = 422 mas (32\,$\lambda/D$ for 600\,nm, 9\,m aperture).
Dynamic contrast is 0.1 of raw contrast. 
The post-processing contrast factor is 0.29 for detection, 
and 0.1 for spectra.
An integration time limit of 60 days was imposed. 
Astrophysical assumptions are the same as in the Standards Team Final Report 
\cite{morgan2019sdetfinal} and are summarized in \ref{sec:astrophysical}. A full description of instrument parameters is in \ref{sec:instrument} and is consistent with parameters used for LUVOIR-B\cite{luvoir2019} and HabEx\cite{habexreport2020}.

The observing scenario is a 5-year mission with 50\% time for exoplanet observations.
The blind-search survey for exo-Earths uses the 500--600 nm band
at $\textrm{SNR} = 7$, and orbit determination by
three photometric observations.
Once a target is promoted for spectral characterization, the targets are observed spectrally as soon as a sufficiently long observing window is available and are prioritized by benefit/cost, ie completeness divided by integration time.Spectral characterization observations 
always have priority over photometric imaging observations.
The exo-Earth is spectrally characterized at quadrature in one of three wavelength passbands. For this study, spectral characterization is performed once at only one of the passbands.


\subsection{Overview of Yield Code}
\label{sec:exosims}
\begin{figure}
   \begin{center}
   \begin{tabular}{c} 
   \includegraphics[width=.93\textwidth]{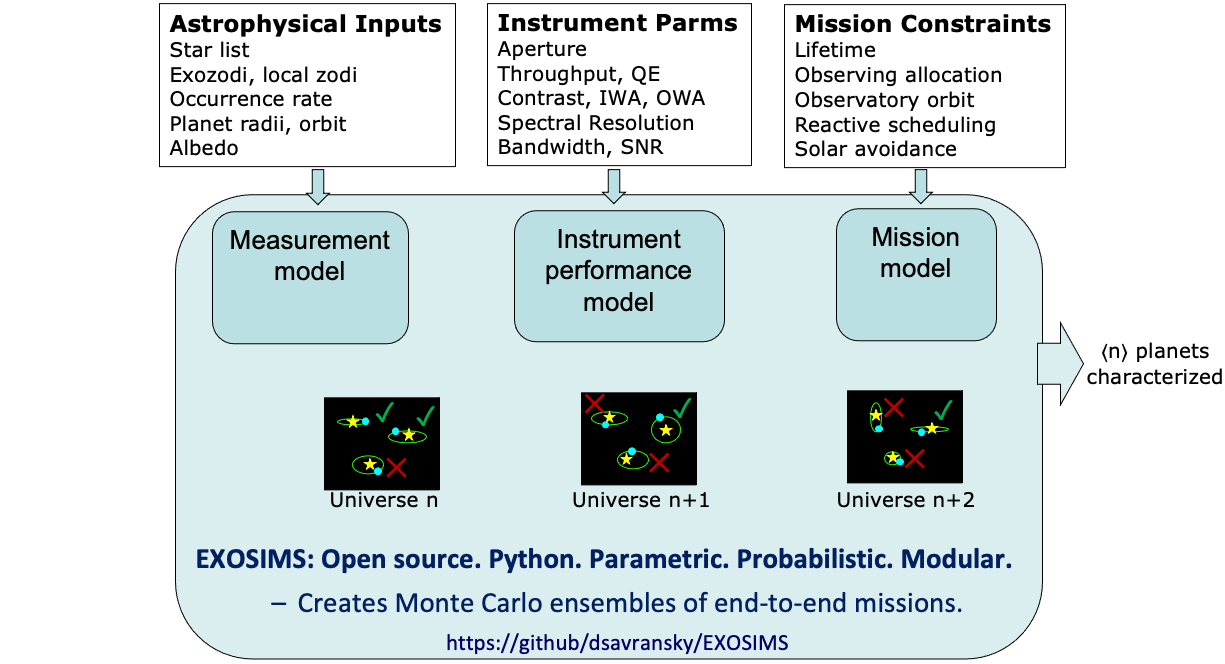}
   \end{tabular}
   \end{center}
   \caption[example] 
   { \label{fig:exosims} 
 EXOSIMS is an open-source, parametric, python-based exoplanet direct-imaging mission simulation code. EXOSIMS creates Monte Carlo ensembles of end-to-end missions with observation scheduling.}
\end{figure}

We used EXOSIMS\cite{exosims-github} to perform all of the yield modeling in this study. EXOSIMS is an open-source, parametric, Python-language exoplanet direct imaging mission simulation code. As illustrated in Figure \ref{fig:exosims}, EXOSIMS includes an astrophysics simulation of synthetic exoplanets and astrophysical noise sources; an instrument performance model to calculate signal, noise, and integration time; and a mission model to simulate the orbital dynamics of the observatory's orbit and the solar system bodies for stray light avoidance.  EXOSIMS creates Monte Carlo ensembles of end-to-end missions with observation scheduling. Statistics can then be performed over the Monte Carlo ensemble.  We provide an overview of EXOSIMS in Appendix \ref{sec:appendix_EXOSIMS}.  More detailed description can be found in Refs.~\citeonline{morgan2021, morgan2019sdetfinal}.

\subsection{Overview of Sensitivity Analysis Software}
\label{sec:Dakota}

We used various capabilities within 
the Dakota software package
for these experiments.
Dakota \cite{Dakota} has been developed
by Sandia National Laboratories for parametric analyses of
complex computer models, enabling design exploration, 
risk analysis, and quantification of margins and uncertainty.
The gridded experimental design in Sec.~\ref{sec:analysis} used 
Dakota's \texttt{multidim\_parameter\_study} capability, and
the alternate design used \texttt{list\_parameter\_study}.
Dakota has the capability to generate low-discrepancy designs
(\texttt{sample\_type low\_discrepancy}) but we generated the above
design offline using SciPy (\texttt{stats.qmc.Sobol}).
The Dakota templating utility (\texttt{pyprepro})
was used to adapt an exemplar EXOSIMS input script
to the specific parameters and observing modes needed.
Runs were executed using Dakota as the top-level driver
for an ensemble run managed by GNU parallel\cite{tange_ole_2021}.


\section{Results}
\label{sec:results}

\subsection{Yields}
\label{sec:yield-est}

As described in Sec.~\ref{sec:exosims}, yield for a given observing
scenario is computed via a Monte Carlo average, across
the $N_e = 100$ ensemble members, of the yields of
individual simulations:
\begin{equation}
    \hat Y = \frac{1}{N_e} \sum\nolimits_{i=1}^{N_e} Y_i
    \label{eq:avg-yield}
\end{equation}
The Monte Carlo randomization is done over many factors:
\begin{itemize}
    \itemsep0pt 
    \item Planet assignments around stars
    \item Planet orbital parameters
    \item Planet physical properties 
    \item Observing conditions (e.g., exozodi)
    \item Observation scheduling
\end{itemize}
Because observation scheduling reacts to all the other factors,
it seems difficult in general to make statements about which
of these randomized factors dominates the observed scatter of 
the ensemble of yields $\{ Y_i \}_{i=1}^{N_e}$ for a given
scenario.
However, it is true that the yields are independent, because 
mission simulations do not interact, implying that the summands
of~\eqref{eq:avg-yield} are statistically independent.
This allows us to find a simple expression for the accuracy 
of our Monte Carlo yield estimate:
\begin{align}
    \var({\hat Y})
    &= 
    \frac{1}{N_e^2} \var \Bigl( \sum_{i=1}^{N_e} Y_i \Bigr)
    = 
    \frac{1}{N_e^2} \sum_{i=1}^{N_e} \var({Y_i})
    = 
    \frac{1}{N_e} \var({Y_1})\quad\textrm{, and thus}
    \nonumber
    \\
    \std({\hat Y})
    &=
    \tfrac{1}{\sqrt{N_e}} \std({Y_1}) \quad.
    \label{eq:std-yield}
\end{align}
Furthermore, because the yield estimate is a sample mean,
the Central Limit Theorem applies to $\hat Y$, implying its
distribution is well-approximated by a Gaussian for moderately large
$N_e$. 
So, within our MCMS framework, the yield estimate $\hat Y$ and
its standard error as computed from \eqref{eq:std-yield} 
form an essentially-complete description of both the yield
and its dispersion.
The dispersion informs both our choice of $N_e = 100$, and
the ultimate accuracy of the yield models we extract from
the set of scenarios in our experimental design.

\begin{figure}
\centering
   \begin{tabular}{cc} 
       \includegraphics[width=.60\textwidth]{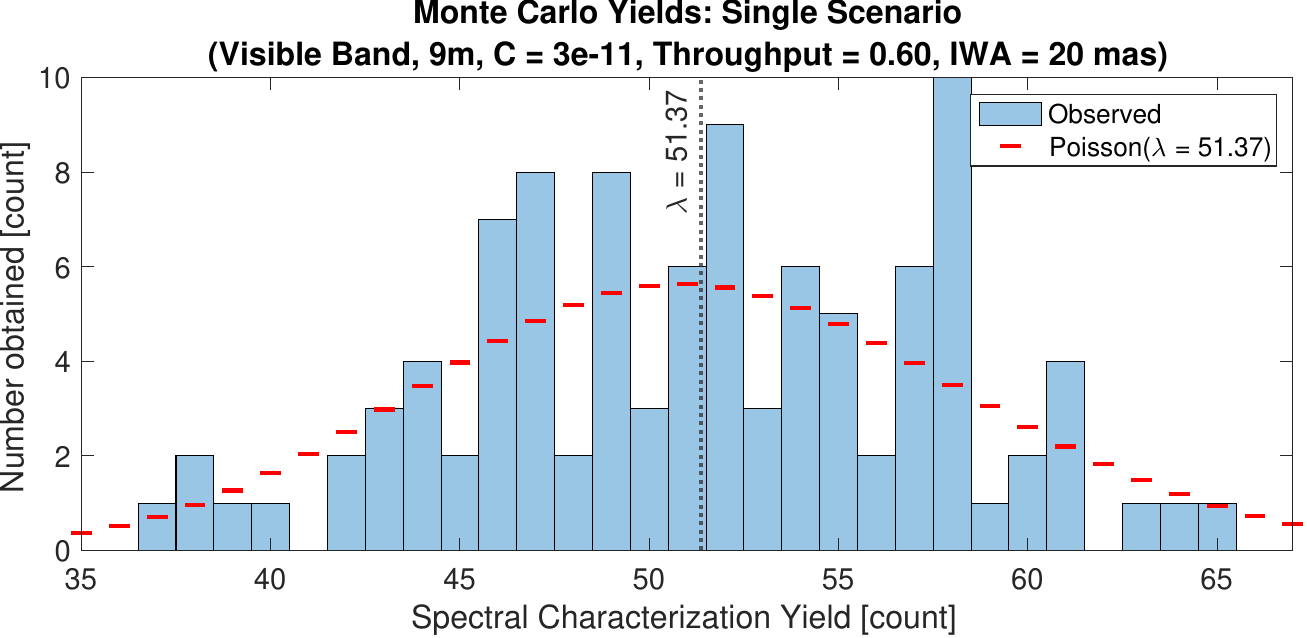}%
       &
       \includegraphics[width=.37\textwidth]{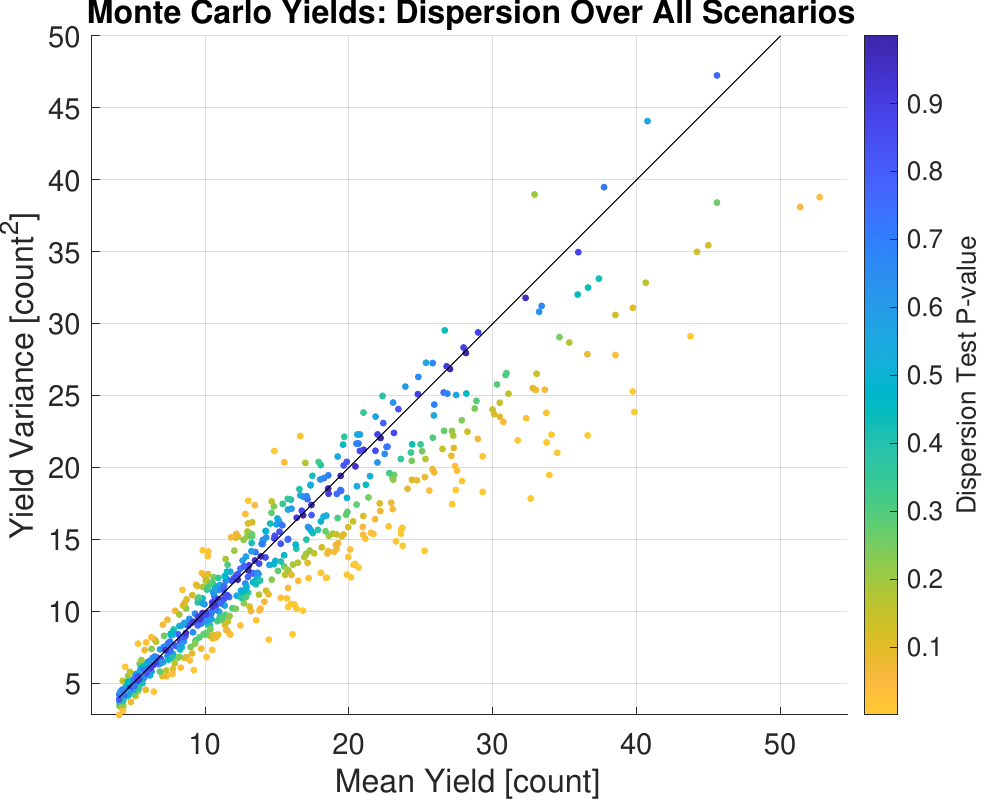}%
   \end{tabular}
   \caption{\label{fig:yield-distribution}%
    Left panel: Observed yield realizations for one scenario, superimposed on the expected number assuming a Poisson distribution of yields.
    Right panel: Scatter plot of observed yield ($\hat Y$) and yield variance
    ($\var(Y)$) for all scenarios. 
    Poisson variables would tend to lie along the mean = variance line (shown
    in black). Points are colored by their p-values according to 
    the Poisson dispersion
    test. About 14\% of scenarios have p-values below 0.05.}
\end{figure}

We can add one more level of detail. 
The yield on any single mission simulation is a sum of 
individual successful spectral characterizations
across many observing attempts. 
For a fixed observing system, success or failure on a single
observation is related to geometric and photometric conditions
at the observing epoch.
Heuristically, a single-mission yield is a sum of many, loosely-coupled, 
small-probability 0/1 events.
Because of this, we might expect Poisson statistics to apply to
the single-run yields $Y_i$ of \eqref{eq:avg-yield}.

\begin{figure}
\centering
    \begin{tabular}{cc} 
       \includegraphics[trim=12mm 0 10mm 0mm,clip,width=.50\textwidth]{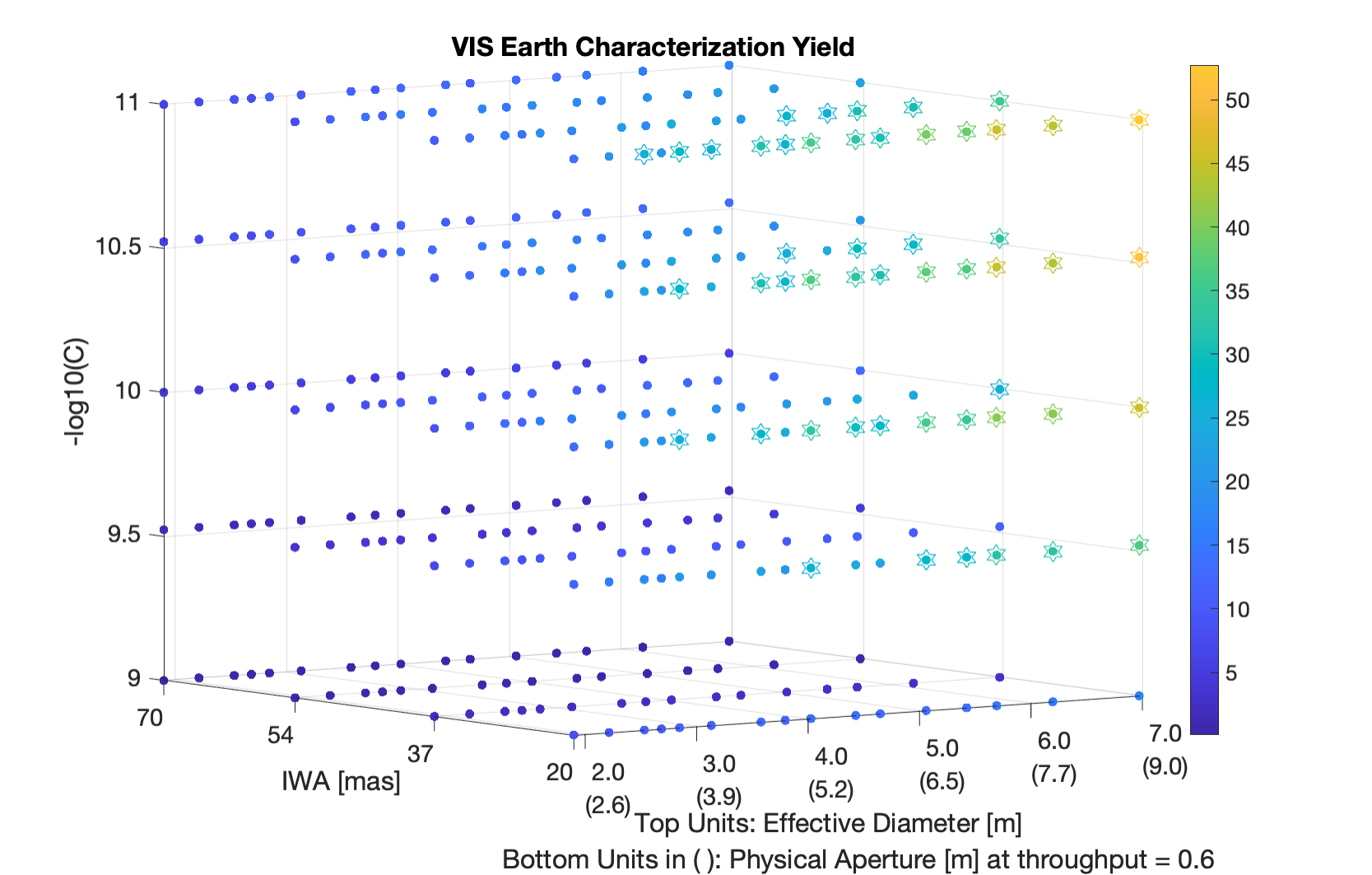}%
       &
       \includegraphics[trim=12mm 0 10mm 0mm,clip,width=.50\textwidth]{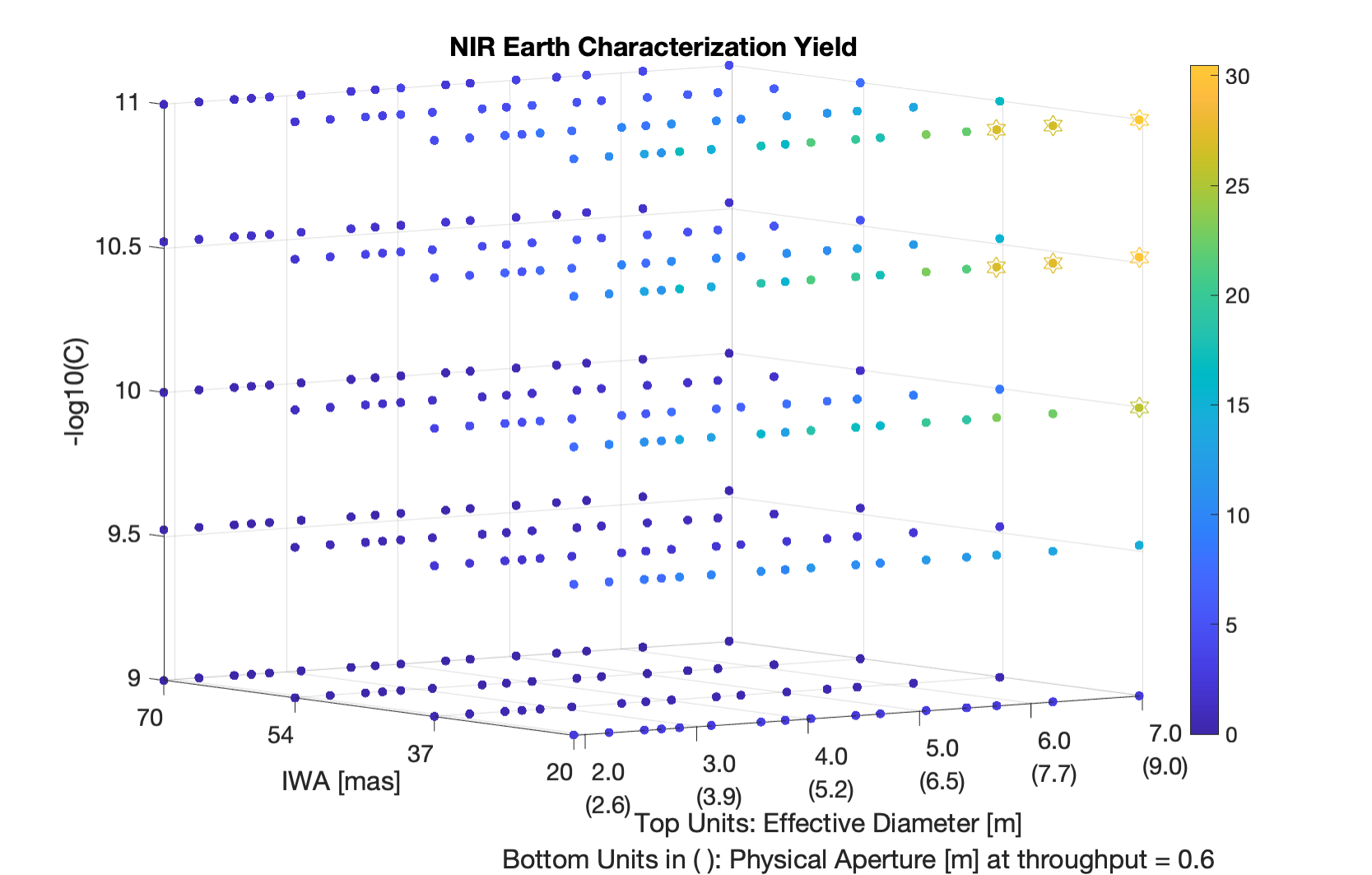}%
       \\
       \multicolumn{2}{c}{\includegraphics[trim=12mm 0 10mm 0mm,clip,width=.50\textwidth]{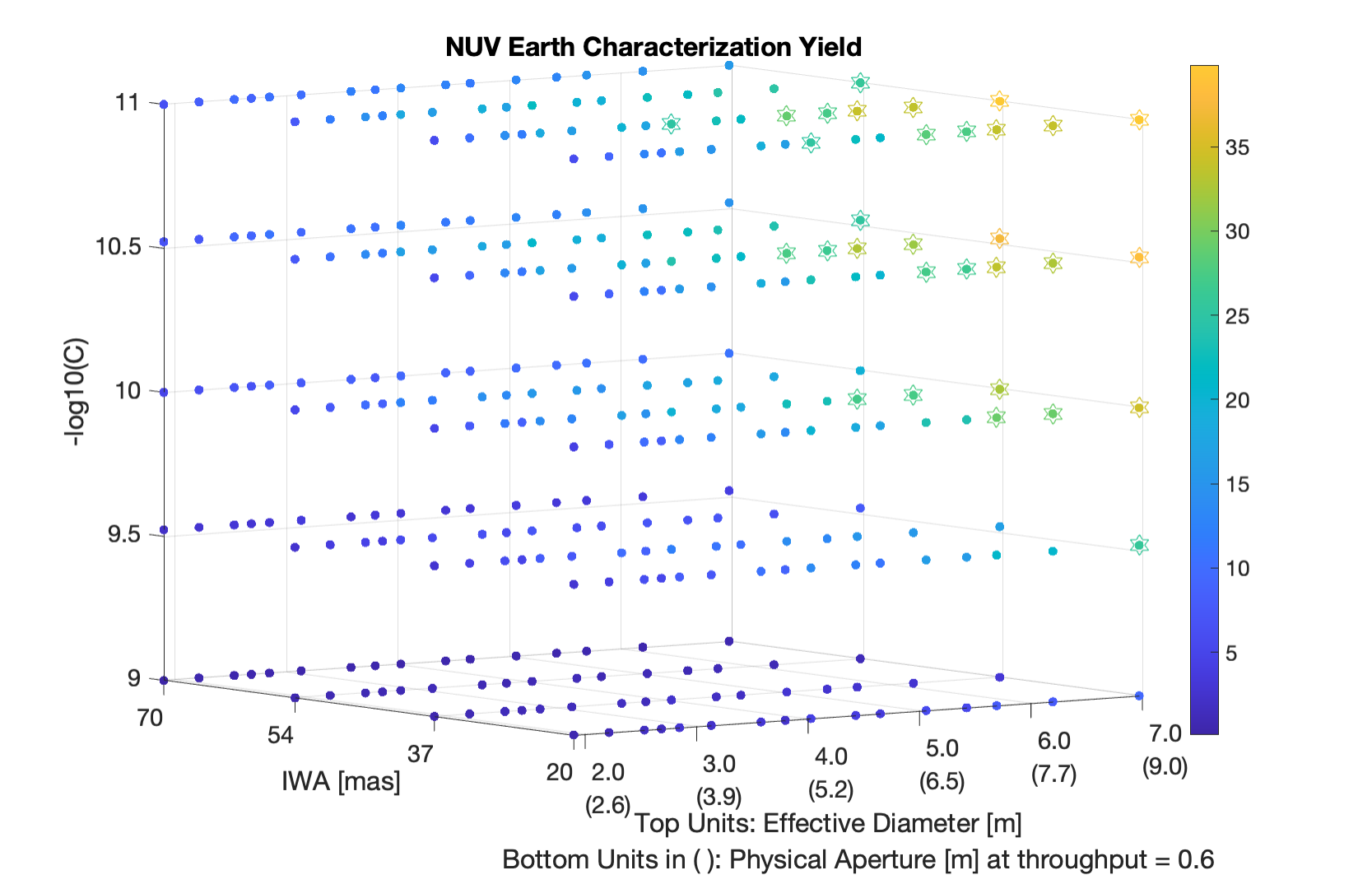}}%
       
    \end{tabular}
    \caption{\label{fig:yield-scatter}%
    Yield estimate $\hat Y$ versus parameters for VIS band scenarios 
    (top left), NIR scenarios (top right), and NUV (bottom).
    Yields of at least 25 are shown with an enclosing star.}
\end{figure} 

Indeed, Figure~\ref{fig:yield-distribution}, left panel,
shows that the histogram of $\{Y_i\}$ for a certain 
scenario overlays well onto the Poisson distribution
with parameter $\lambda = \hat Y$.
The right panel of that figure aggregates over all observing
scenarios with $\hat Y \ge 4$, showing
a scatter plot of the sample mean $\hat Y$ against the 
sample variance computed from the same observations. 
Under a Poisson distribution, these points would be expected to 
lie on the $\textrm{variance} = \textrm{mean}$ line. 
Points are shaded according to the
p-value of the Poisson dispersion test\cite{brown2002poisson}
which examines the test statistic 
\begin{equation*}
    D = (N_s - 1) \var({Y}) / \hat Y
\end{equation*}
under a $\chi^2$ distribution with $N_s-1$ degrees of freedom.
A small p-value indicates departure from Poisson statistics.
In general, we find support for approximate Poisson-like behavior,
but we observe some under-dispersion (points below the
$45^\circ$ line).
This under-dispersion can be explained by a 
coupling between some observations in a given mission simulation.
For instance, some stars with planets present may require 
large integration times, implying multiple visits 
(detection and follow-up characterization) that displace other targets.
Note that these departures from Poisson behavior of 
mission-by-mission yield values do not imply 
that the sample average $\hat Y$ departs 
from approximate Gaussianity.

\subsection{Overall Yield Behavior Relevant to HWO}

Scatter plots of yield versus mission parameters
(Fig.~\ref{fig:yield-scatter}) give a good overall picture
of the yield.
In these plots, we define an effective area and effective diameter:
\begin{equation}
    \Aeff = \Deff^2 = \tputWord \cdot \textrm{Diameter}^2 = \tput \diam^2
    \label{eq:eff-area}
\end{equation}
which allow us to compress the four continuous-valued variables of 
our parametric study into three variables for visualization.
The choice of values is such that there is no overplotting of points.

In both plots, we see yield clearly increase with decreasing IWA
and increasing contrast (more targets available), 
and typically increase with $\Deff$ (more photons).
The yield with respect to the combined parameter
$\Deff$ is largely, but not completely monotonic.
All three plots show a significant ``top-right'' zone of
performance with $\hat Y \ge 25$ 
that responds to HWO mission requirements.
In NIR, we see a drop in yield, relative to VIS,
due to the working angle
limitation at longer wavelength. In the NUV, the yield is lower then in the visible, larger due to diminished stellar flux resulting in longer integration times and targets lost to the integration time cutoff policy.

\begin{figure}
\centering
       \includegraphics[trim=90mm 30mm 90mm 30mm,clip,width=.95\textwidth]{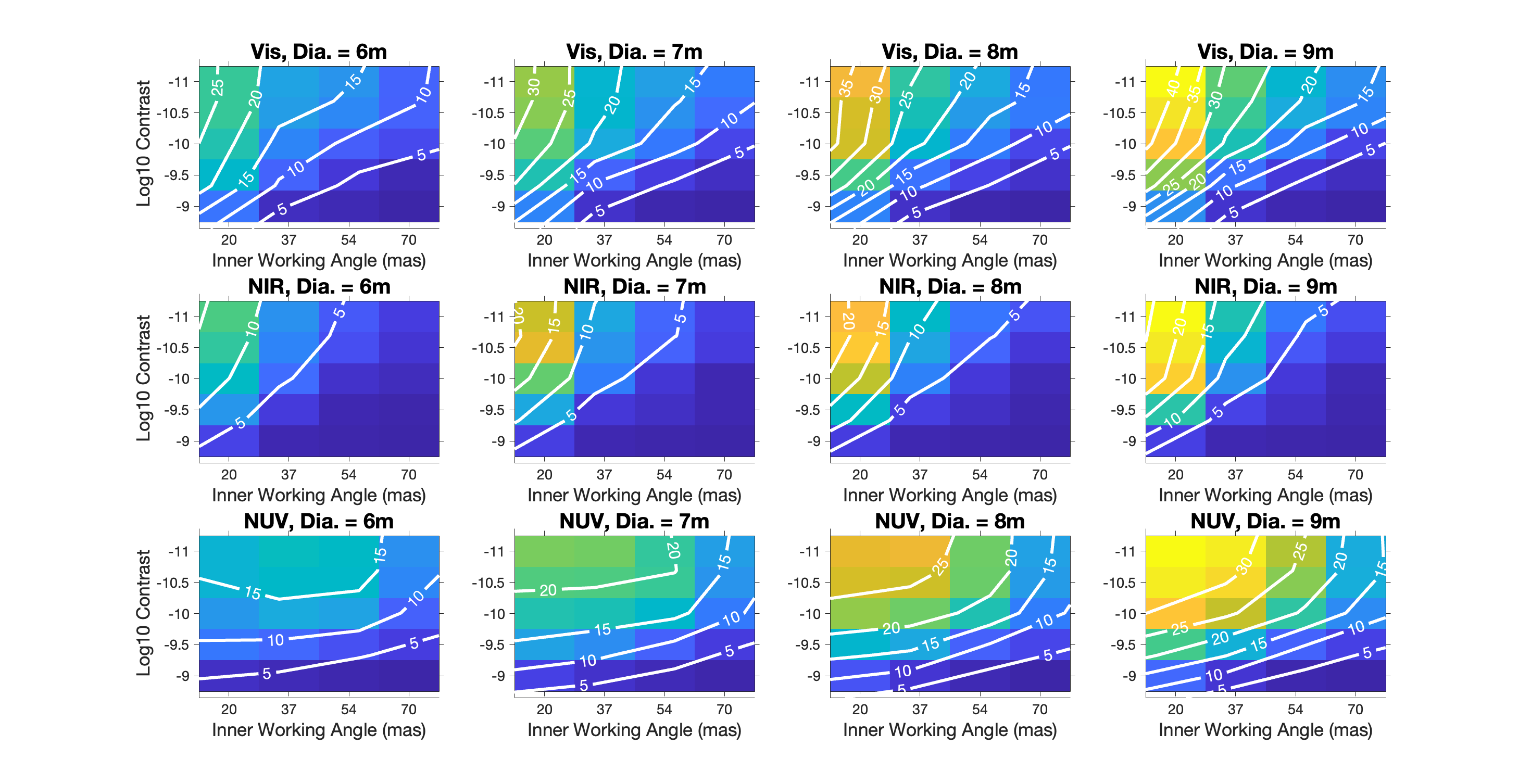}%
    \caption{\label{fig:IWAcontrast_heatmaps}%
    Yield data in the IWA vs Contrast plane for each of the sampled diameters (columns) and each of the wavelength passbands (rows). White contour lines show yield for every 5 exo-Earths.}
\end{figure} 

Figure \ref{fig:IWAcontrast_heatmaps} shows cuts through the yield data in the plane of IWA vs. contrast for each of the diameters at each of the wavelength passbands. The interaction between IWA and contrast is easily seen; the orientation of the gradient varies with both aperture diameter and wavelength and the strength of the gradient has a strong dependence on wavelength. In the NIR, this is most dramatically seen; yield has very sensitive to IWA, and sensitivity to contrast is seen only at small IWA and large apertures, particularly because of low yields. the geometric challenge of the NIR 1.65 $\mu m$ can be clearly seen. The NIR achieves 25 exo-Earths characterized only at 9-m aperture and 20 mas IWA, which corresponds to $1\, \lambda /D$.

\subsection{Yield Behavior Local to a 6m HWO}
\begin{figure}
\centering
    \begin{tabular}{cc} 
       \includegraphics[trim=12mm 0 10mm 0mm,clip,width=.40\textwidth]{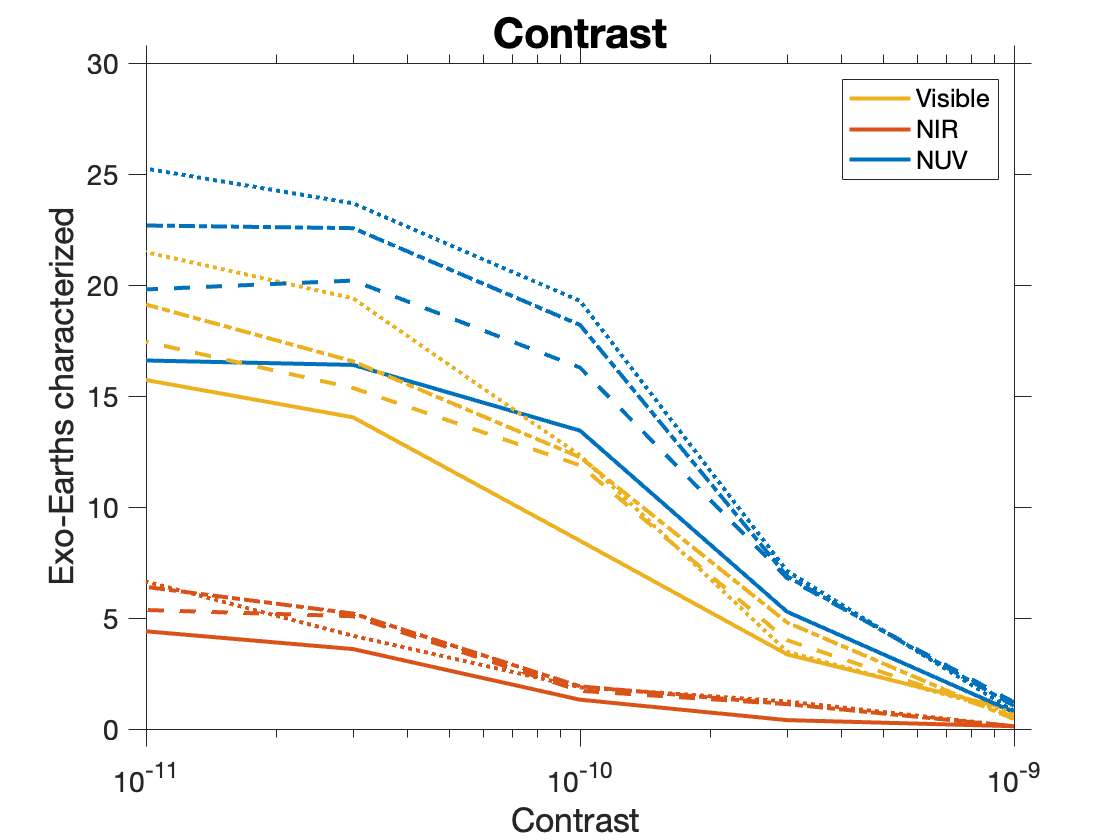}%
       &
       \includegraphics[trim=12mm 0 10mm 0mm,clip,width=.40\textwidth]{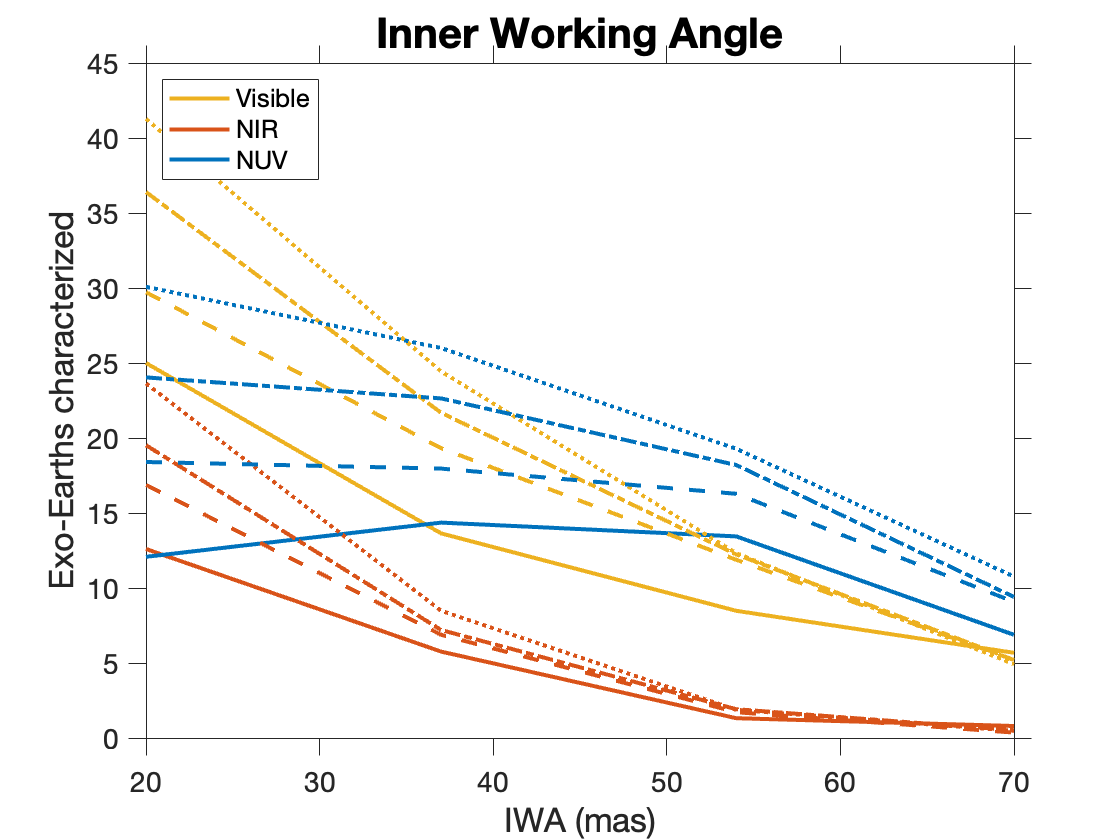}\\
       \includegraphics[trim=12mm 0 10mm 0mm,clip,width=.40\textwidth]{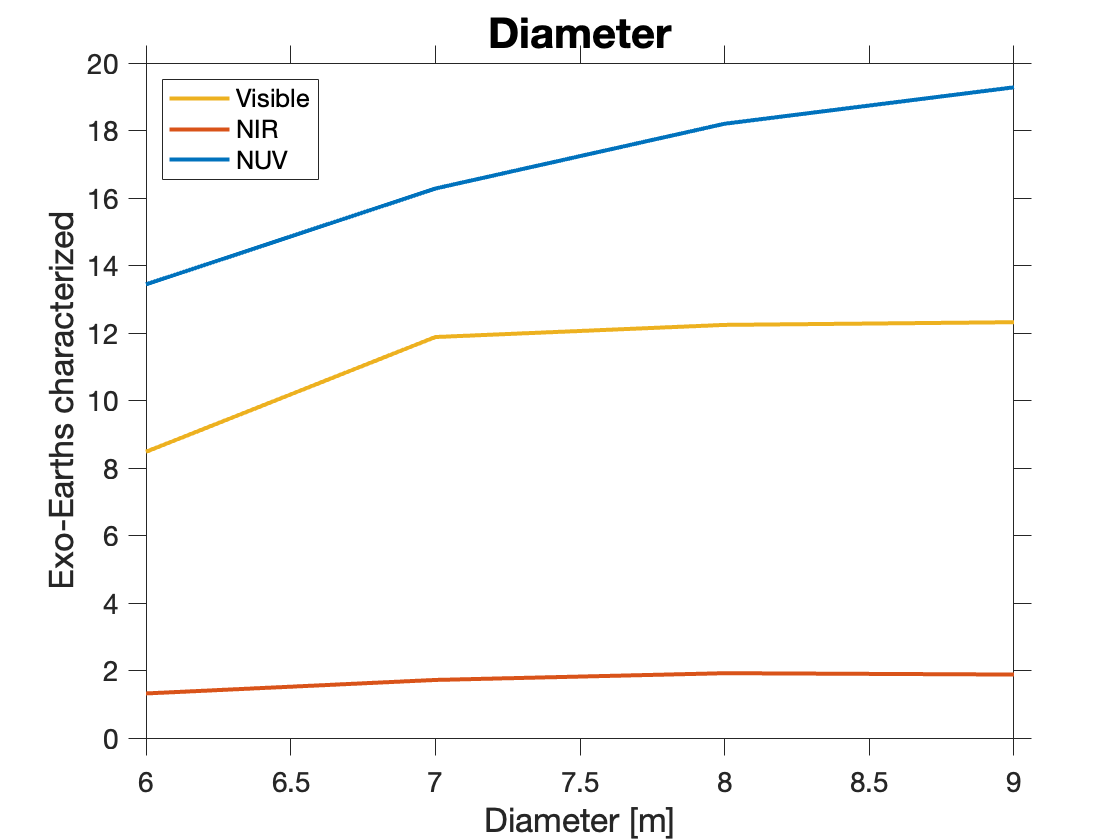}%
       &
       \includegraphics[trim=12mm 0 10mm 0mm,clip,width=.40\textwidth]{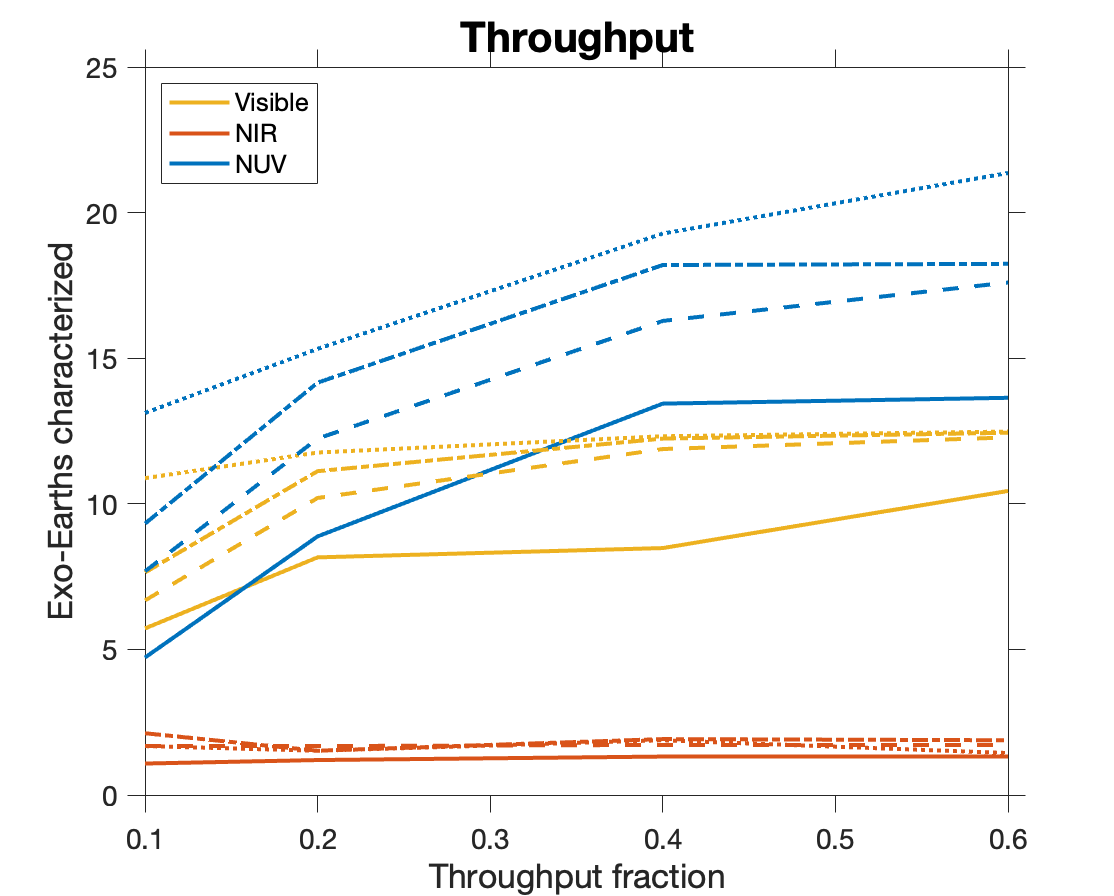}
    \end{tabular}
    \caption{\label{fig:local}%
    Sections through the yield response function around a nominal HWO design. As in Figure \ref{fig:LvDist}, yellow lines denote the visible passband, red line the NIR, and blue lines the NUV. The aperture sizes are indicated by: 6m solid line, 7m dashed line, 8m dash-dot line, 9m dotted line.}
\end{figure}

Figure \ref{fig:local} shows the yield behavior for each of the parameter sweeps centered on a the sample point closest to a nominal HWO design: 6m diameter, 1E-10 contrast, 0.4 throughput, and 70 milliarcsecond IWA. Contrast has a very low yield at $10^{-9}$ for all wavelengths and aperture sizes. As contrast improves, the sensitivity to aperture size and wavelength increases, with the exception of NIR, which has a low yield in the space local to a 6m HWO over all the parameters studied. IWA shows a week sensitivity to aperture diameter at 70 mas, and increased sensitivity to both aperture diameter and wavelength with improving IWA. NUV yields have diminished returns at 20 mas compared to 37 mas, likely because the increased number of NUV targets accessible by NUV at 20 mas are not also accessible by the visible coronagraph at 575 nm for the blind-search survey and thus are not discovered.

The trend in diameter is largely due to collecting area, since the IWA is defined in arcseconds and not $\lambda / D$. Interestingly, in the visible the diameter saturates above 7 m for the other parameter values at this local space. In the NUV, yield increases with the diameter, again demonstrating the importance of gathering photons in  the face of diminishing stellar flux.  The throughput follow similar trends for all aperture sizes, underscoring the importance of photon gathering for the NUV and that in the visible a design of 7m aperture, 0.4 throughput, $1^{-10}$ contrast, and 70 mas IWA has sufficient photons for the water search spectra quality.

\section{Sensitivity Analysis}
\label{sec:analysis}

We performed a sensitivity analysis of the yield to
the instrumentation parameters to understand overall
qualitative behavior of yield, to investigate parameter
interactions, and as a feasibility study towards 
developing a parametric yield model for coronagraph designs.
We examined two overall approaches: response surface models
using polynomial fits to yield, and a 
nonparametric yield model using random forests.

We view the (unobserved, idealized) yield as a function of inputs:
\begin{equation}
    Y = f(\textrm{mode}, \diam, \contrast, \tput, \iwa) = f(\theta)
\end{equation}
where $\theta$ generically denotes 
a tuple of design parameters.
The idealized yield is viewed as an expectation value, 
integrating over the distributions of the many sources of uncertainty
mentioned in Sec.~\ref{sec:yield-est}.

We have two experimental datasets, which are input-output
characteristics over the experimental designs:
\begin{align*}
    \mathcal{D}_\grid &= 
    \{ (\theta,\, \hat Y(\theta)) \suchthat \theta \in \mathcal E_\grid \}
    \\
    \mathcal{D}_\sobol  &= 
    \{ (\theta,\, \hat Y(\theta)) \suchthat \theta \in \mathcal E_\sobol \}
    \quad.
\end{align*}
As noted, the Central Limit Theorem makes it reasonable to assume that 
\begin{equation}
    \hat Y(\theta) = f(\theta) + \epsilon_\theta
\end{equation}
where all $\epsilon_\theta$ are independent, zero-mean Gaussian with 
variance determined as in Sec.~\ref{sec:yield-est}.

\subsection{Response Surface Modeling}
\label{sec:surface}

\begin{table}
    \caption{Quality of Yield Response Surface Across Bands and Datasets%
    \label{tab:rsm_stats}}
    \centering
    \begin{tabular}{clccccccc}
        \hline\hline
    \textbf{Band} 
      & \textbf{Model} 
      & \multicolumn{3}{c}{\bf Fitting: Pooled} 
      & \multicolumn{2}{c}{\bf Grid} 
      & \multicolumn{2}{c}{\bf Sobol} 
    \\
    ~ & \textbf{Form} & $R^2$ & RMS & MAE & RMS & MAE & RMS & MAE
    \\
    \hline\hline
     NIR & Linear & 
    	0.851 & 1.973 & 1.577
    	 & 2.241 & 1.845
    	 & 1.572 & 1.240
     \\ 
        \texttt{"} & Interactions & 
    	0.974 & 0.822 & 0.658
    	 & 0.861 & 0.714
    	 & 0.769 & 0.588
     \\ 
     VIS & Linear & 
    	0.837 & 3.467 & 2.589
    	 & 3.958 & 3.043
    	 & 2.871 & 2.116
     \\ 
     \texttt{"} & Interactions & 
    	0.975 & 1.352 & 1.066
    	 & 1.488 & 1.200
    	 & 1.195 & 0.928
     \\ 
     NUV & Linear & 
    	0.730 & 3.679 & 2.911
    	 & 4.049 & 3.172
    	 & 3.302 & 2.671
     \\ 
     \texttt{"} & Interactions & 
    	0.946 & 1.649 & 1.285
    	 & 1.640 & 1.246
    	 & 1.658 & 1.321 
      \\
    \hline
    \end{tabular}
\end{table}

Conventional response surface modeling \cite{BD2007} is
a statistical approach to understanding a sparsely-sampled\pagebreak[4]
function.
The response surface methodology fits a parametric function
to these datasets. Using a squared-error loss which is
appropriate for Gaussian errors, we can use least
squares to obtain parametric models.
It is helpful to use a variable transformation on 
contrast: our polynomial fit is expressed in terms
of $\tilde \contrast = \log_{10} \contrast$.
The exploratory fits below used the gridded data $\mathcal D_\grid$;
the final fitted model used the pooled data from both 
experiments.
We fitted separate models for VIS, NIR, and NUV, and we 
removed all parameters $\theta$ for which 
$\hat Y(\theta) < 5$.
We are not typically interested in these parameters,
and we did not want to adapt the fitted coefficients
to model such low yields.

The univariate linear response surface models, when fitted
on the gridded experiment $\mathcal D_\grid$, and centered about 
a nominal design point 
\begin{equation*}
    \theta_0 = 
    (6\,\textrm{m},\, 10^{-10},\, 50\,\textrm{mas},\, 0.40)
    \quad,
\end{equation*}
are
%
\begin{align}
f_1^{\textrm{NIR}}(\theta) &= -5.31 
        + 2.65 \, (\diam - 6)
        + 5.59 \, (\tilde \contrast - 10)
        + 16.91 \, (\tput - 0.4)
        - 0.57 \, (\iwa - 50)
    \\
f_1^{\textrm{VIS}}(\theta) &= \phantom{-} 8.28 
        + 3.07 \, (\diam - 6)
        + 8.86 \, (\tilde \contrast - 10)
        + 20.39 \, (\tput - 0.4)
        - 0.45 \, (\iwa - 50)
    \\
f_1^{\textrm{NUV}}(\theta) &= \phantom{-} 9.70 
        + 3.19 \, (\diam - 6)
        + 7.91 \, (\tilde \contrast - 10)
        + 22.91 \, (\tput - 0.4)
        - 0.20 \, (\iwa - 50)
\quad.
\end{align}
All coefficients had p-values below $10^{-20}$, indicating
strong support for the term being required for the fit.
Focusing on VIS, and defining the residual
$\epsilon_1(\theta) = \hat Y(\theta) - f_1^\textrm{VIS}(\theta)$, we 
compute the residual RMSE (of $\epsilon_1$) is 4.01, 
and the coefficient of determination $R^2 = 0.82$.
Using \eqref{eq:std-yield}, we find
the Monte Carlo sampling error of $\hat Y$ is
typically about 0.45, so this model RMSE indicates 
significant lack of fit.
When viewed in parameter space, 
the residuals show some smooth structure,
also indicating lack of fit.
Some trends still emerge from these models, including 
low yield and sensitivity to IWA for NIR, and diminished
sensitivity to IWA for NUV.

\begin{figure}
\centering
    \begin{tabular}{cc} 
       \includegraphics[trim=0mm 0 0mm 0mm,clip,width=.40\textwidth]{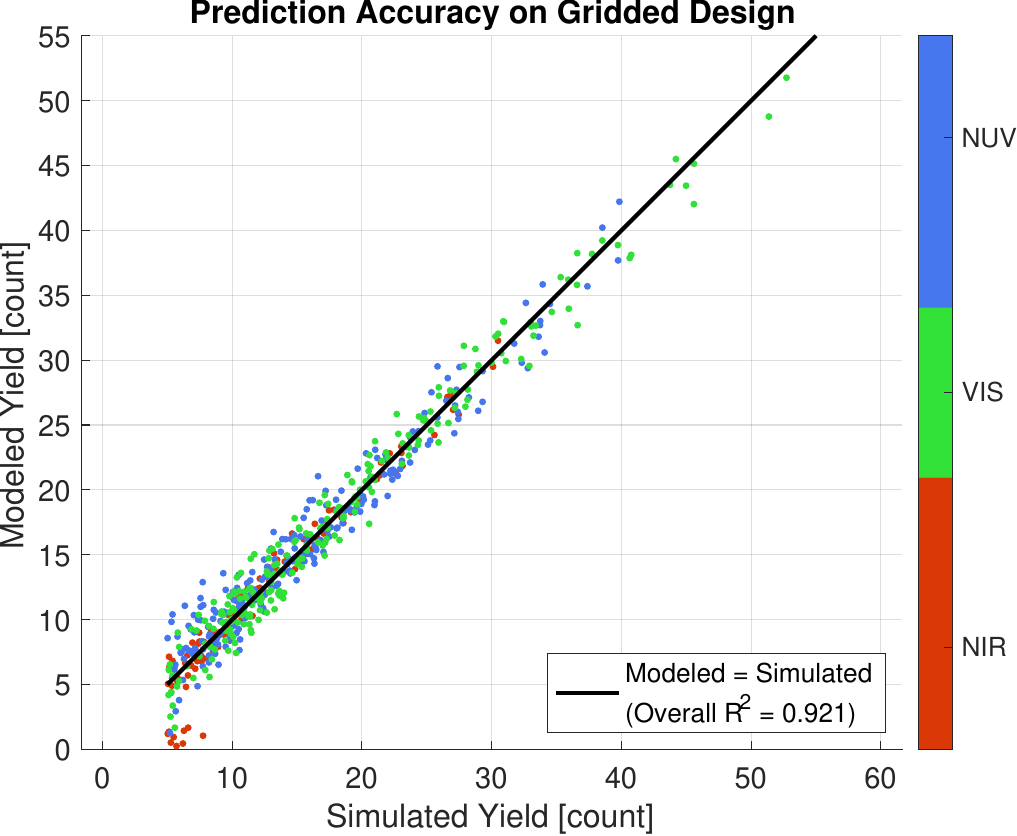}%
       &
       \includegraphics[trim=0mm 0 0mm 0mm,clip,width=.40\textwidth]{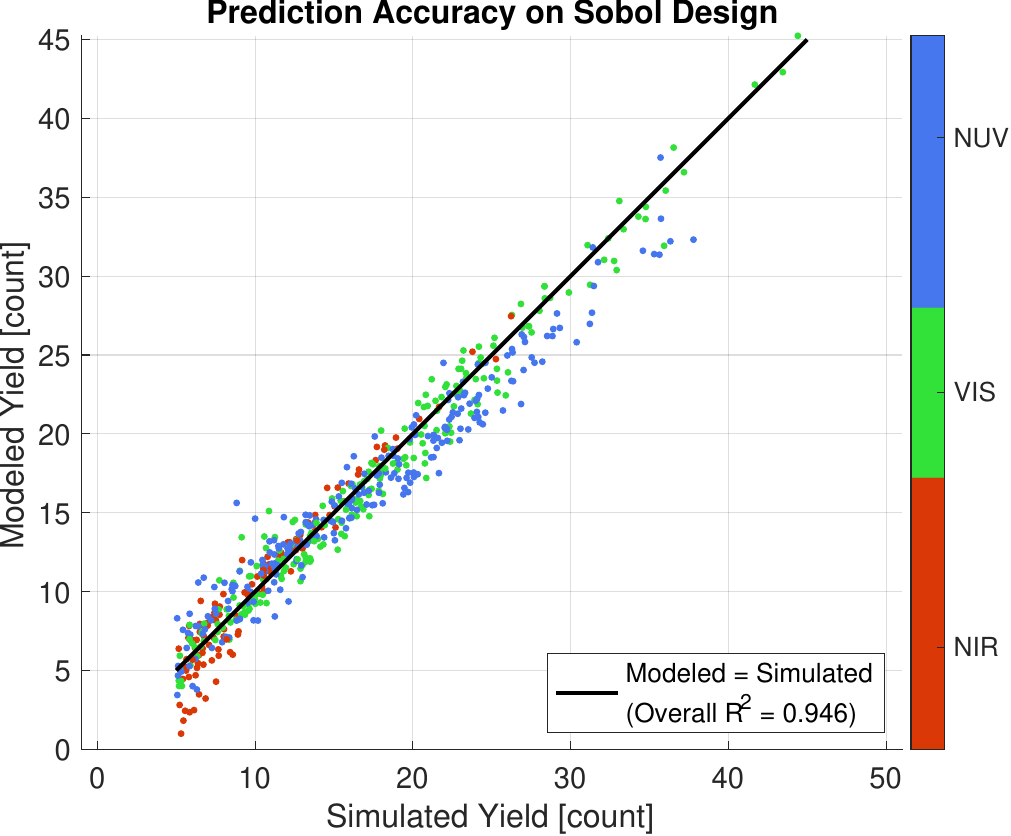}
    \end{tabular}
    \caption{\label{fig:yield-predict}%
    Yield prediction accuracy for the second-order
    response surface model, with interactions: left panel, 
    on training data $\mathcal D_\grid$; right
    panel, on held-out data $\mathcal D_\sobol$.
    In general, we see yield predictions consistent
    with a residual RMSE of 1--2 characterizations.}
\end{figure}

Parameter interactions could improve the fit, 
so we introduce second-order interaction terms (parameter cross-products)
and fit a second model for VIS yield, centered about the nominal point:
\begin{equation}
\setlength\arraycolsep{2pt}
f_2(\theta) = 11.09 
      + [ 1.20 ,\, 10.67 ,\, 5.83 ,\, -0.33  ] 
      \cdot (\theta - \theta_0)
      + (\theta - \theta_0)' 
      \begin{bmatrix*}[r]
         0.00 & 2.15 &   0.00 & -0.10 \\
         0~~ & -7.01 &  13.86 & 0.11 \\
         0~~ & 0 ~~~ & -27.09 & -0.64 \\
         0~~ & 0 ~~~ & 0 ~~~ & 0.01 \\
      \end{bmatrix*}
      (\theta - \theta_0)
\label{eq:rsm-interact}
\end{equation}
The residuals $\epsilon_2(\theta)$ have 
RMSE of 1.49, and the model has $R^2 = 0.976$, which are both  
much more favorable.
In performing the least-squares fit, we found that all interactions, 
except between diameter and throughput,
have significant support according to their p-values, so we removed
that one term from the final fit.
We used the same strategy of pruning model terms with lack of support
for the other two bands.
Figure~\ref{fig:yield-predict} shows a satisfying fit quality,
consistent with these RMSE's, for the models
when fitted on $\mathcal D_\grid$ and evaluated on $\mathcal D_\sobol$.

\begin{figure}
\centering
    \begin{tabular}{@{}c@{}c@{}c@{}}%
       \includegraphics[trim=8mm 0 8mm 0mm,clip,width=.33\textwidth]{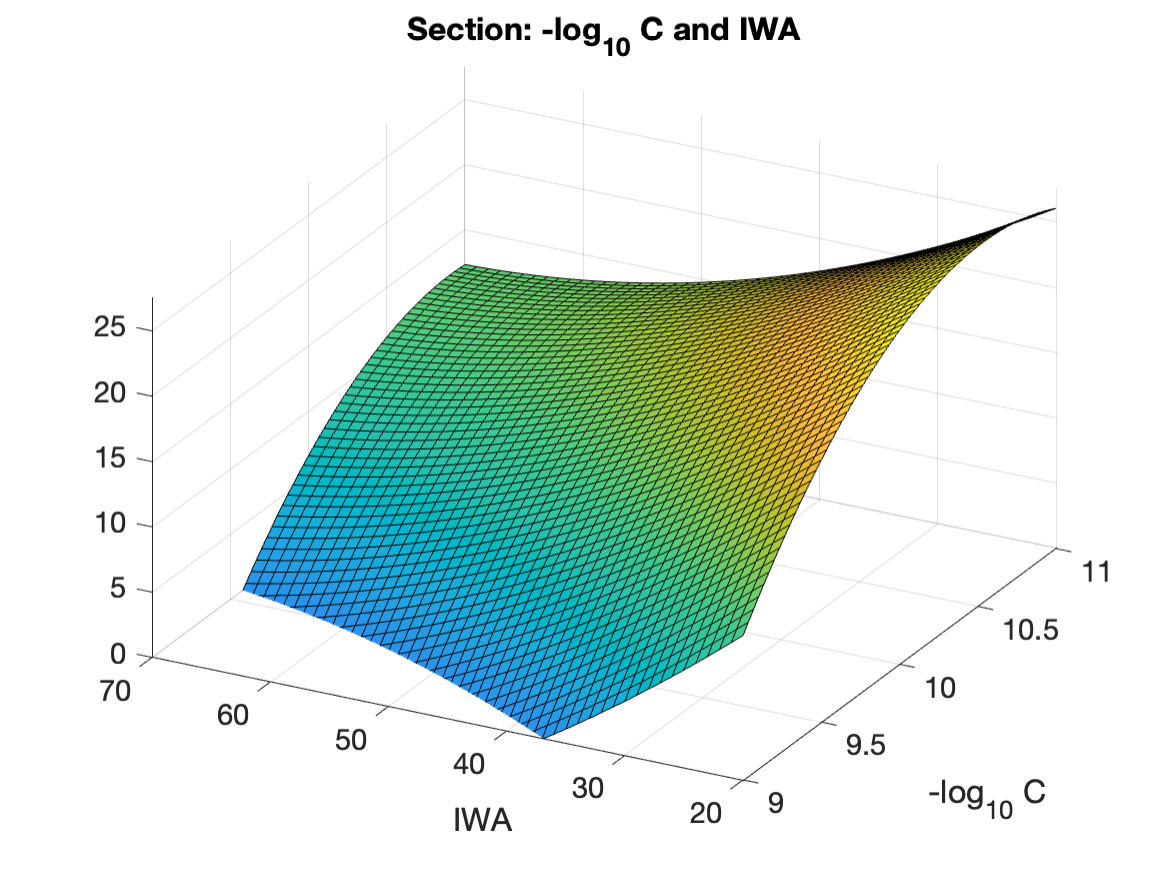}%
       &%
       \includegraphics[trim=8mm 0 8mm 0mm,clip,width=.33\textwidth]{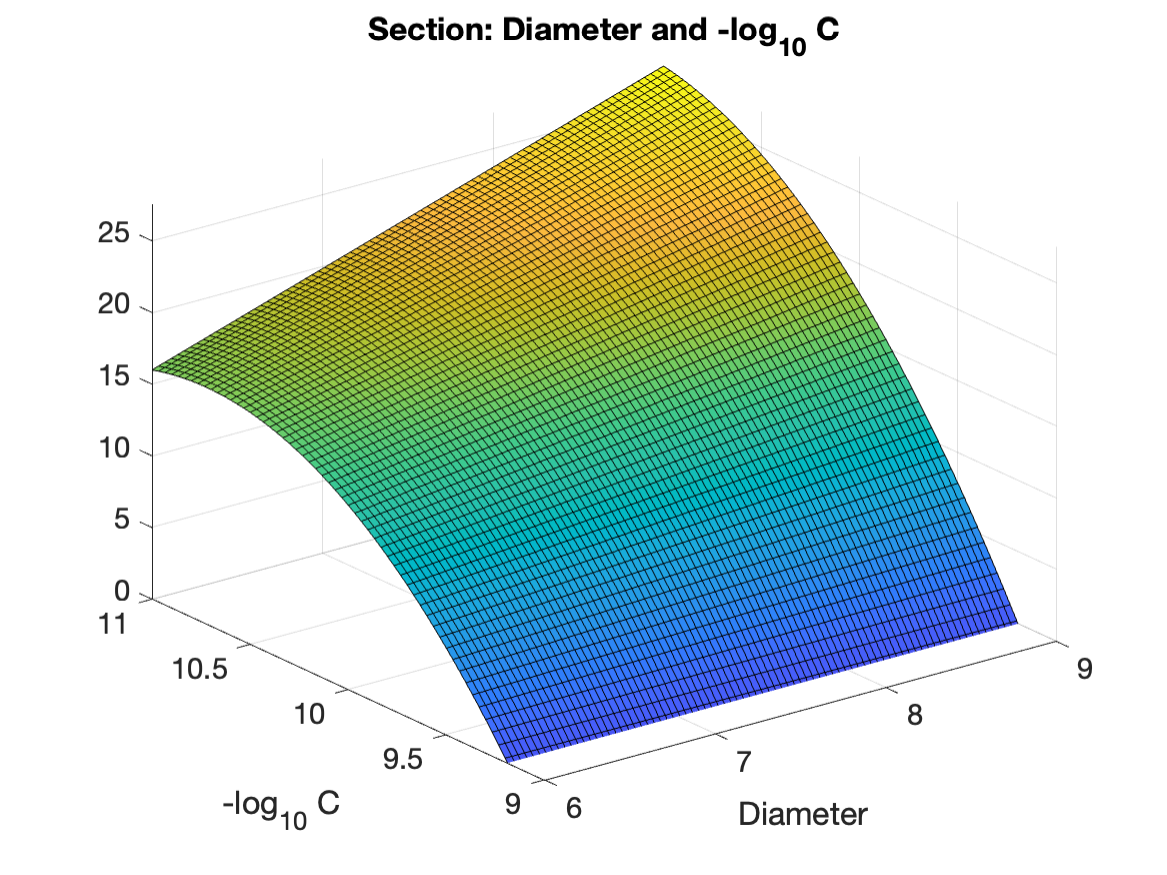}%
       &%
       \includegraphics[trim=8mm 0 8mm 0mm,clip,width=.33\textwidth]{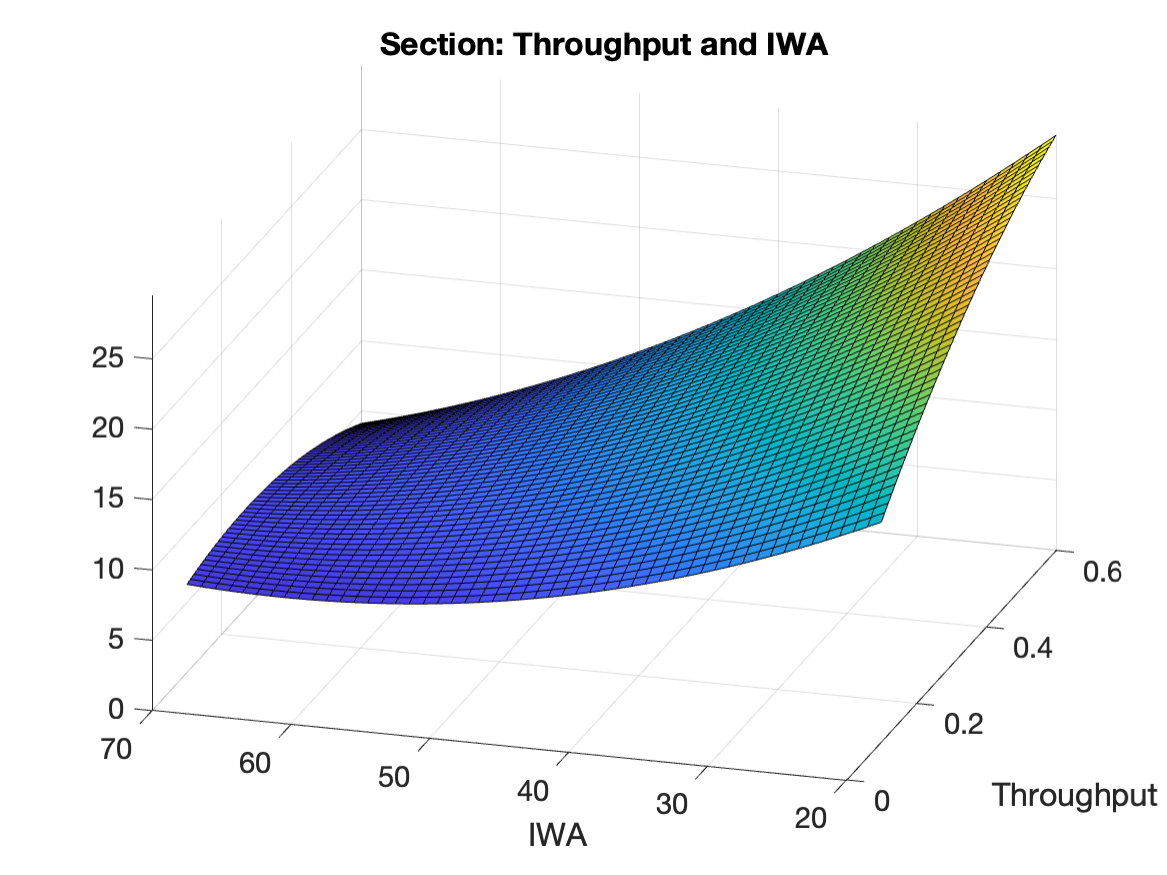}%
    \end{tabular}
    \caption{\label{fig:yield-sections}%
    Sections through the second-order
    response surface model about a nominal design point.
    Left: As a function of $(\tilde C,\, \iwa)$, showing the large influence of 
    these two parameters. 
    Center: as a function of $(\diam,\, \tilde \contrast)$,
    showing  greater sensitivity to diameter 
    at favorable contrast ($10^{-11}$) than
    at worse contrast ($10^{-9}$).
    Right: as a function of $(\tput,\, \iwa)$, showing a greater sensitivity 
    to throughput at low IWA.}
\end{figure}

We can use these second-order models, which have acceptably-low error
provided the yield exceeds 5, 
to understand the contours of the yield surface parametrically.
To do this, we fit the model containing interaction terms
to the pooled dataset. The fit of these models is summarized
in Table~\ref{tab:rsm_stats}. In terms of mean absolute error,
the VIS band has error averaging 1 characterization over the
parameter domain of interest.
We can then evaluate this model at any set of values in the 4D space
of interest, such as a dense grid.
Fig.~\ref{fig:yield-sections}, shows VIS yield
behavior around a nominal design point
\begin{equation*}
    \theta_0 = 
    (\textrm{VIS}, 6\,\textrm{m},\, 10^{-10},\, 50\,\textrm{mas},\, 0.40)
\end{equation*}
by viewing the 2D sections through the parametric 
yield surface at that point.
We show three of the set of six ($4\cdot 3 / 2$) sections.
The interaction terms of~\eqref{eq:rsm-interact} have an evident 
influence in all the plots.
For instance, in the rightmost plot, we see greater sensitivity
of yield to throughput at low IWA.
One explanation is that 
yield is generally very poor for instruments only
capable of high IWA, 
but as the instrument IWA shrinks towards 20\,mas, 
more targets become available, and throughput becomes
a limiting factor on yield.

\subsection{Random Forest}
\label{sec:forest}

Random forest\cite{beiman2001rf} (RF) is an ensemble learning method widely used for classification and regression tasks in various scientific fields. This method operates by constructing multiple decision trees during training and outputting the mode of the classes (classification) or mean prediction (regression) of the individual trees. In this work, we used the RF to learn and interpret the correlations within the dataset we calculated. Using RF to model the response surface $f(\theta)$ can eliminate the potential for lack-of-fit of a parametric model. We decided to use a random forest over other potential machine learning methods as it presents key advantages that benefits this work such as: 1. when using RF, little data preprocessing is needed (e.g., no normalization or standardization required), 2. overfitting issues are avoided when using enough trees in the ensemble; adding more trees might decrease the error as the RF output is the mean prediction of all the trees in the ensamble, and 3. RF is one of the most interpretable machine learning models.

The first step we took was to filter our dataset for yields that are lower than 5. This is because we are not interested in yield properties (such as parameter interactions) at very low yields. It is intuitive that bad combinations of certain parameters (e.g. contrast worse than 10$^{-9.5}$, or small diameter or throughput) result in low yields. Once the dataset has been filtered we also divided it into the three different wavelength bands. Therefore, out of the 960 simulations of the original grid, we have 233 samples in the VIS, 233 in the NUV, and 118 in the NIR. The lower number of samples in the NIR is due to 
filtering of low-yield parameters.
Before trees are selected, we added a supplemental column into the dataset, combining diameter and throughput: the effective area $\Aeff$ of \eqref{eq:eff-area}.
Our initial dataset is therefore comprised of five input features, i.e., Diameter, Throughput, Effective Area, IWA, and Log(contrast), and one output feature, i.e., Earth-like\pagebreak[4] characterized planets (yield). As usually done with machine learning algorithms, the dataset is split into \textit{training} and \textit{test} sets in order to fit and evaluate the performance of the model. 
For the RF, we used the pooled dataset 
$\mathcal D_\pool = \mathcal D_\grid \cup \mathcal D_\sobol$ 
(see Section \ref{sec:analysis}). 
The total dataset has then been randomly divided into 90\% training and 10\% testing.

Two main hyper-parameters of the RF need to be tweaked before we can  interpret the correlations in our dataset. The first one is the number of trees in the ensemble and the second one is the number of splits per tree of the ensemble. The first parameter is intuitive as it defines the total number of the predictions that will be used for the output mean prediction, the second parameter define how many time the dataset is split at each node based on the best feature from a random subset of features of the dataset. We tested several trees/splits configurations for our RF and we ultimately adopted 100 as the number of trees and 7 as the maximum depth. This gives us a good configuration for model speed and precision. Table \ref{tab:rf_stats} reports the statistics of the RF. We calculated the mean absolute error (MAE) on the training and test sets, as well as the out-of-bag (OOB) error. The latter is a method of measuring the prediction error of a machine learning models utilizing bootstrap aggregating (bagging), as does the RF. In this case, for each of the single sample to evaluate, we only use those trees that did not split on the specific sample, or formally, OOB error is the mean prediction error on each training sample $x_i$, using only the trees that did not have $x_i$ in their bootstrap sample. This is a metric that measures the generalization of the model: the closer the OOB error (an error on the training) is to the testing MAE, the more general the model is. By looking Table \ref{tab:rf_stats}, we underline that for each of the three wavelength bands, the RF using 100 trees and 7 maximum splits per tree is well generalized on the dataset.

\begin{table}[h!]
    \caption{Random forest training and evaluation statistics. \label{tab:rf_stats}}
    \centering
    \begin{tabular*}{0.7\textwidth}{@{\extracolsep\fill}lcccc}\hline\hline
    \textbf{Band} & \textbf{MAE on training} & \textbf{MAE on testing} & \textbf{OOB error}\\
    \hline\hline
        VIS & 0.704 & 1.392 & 1.372\\
        NIR & 0.426 & 0.942 & 0.975\\
        NUV & 0.640 & 1.232 & 1.205\\
    \hline
    \end{tabular*}
\end{table}

\begin{figure}
   \begin{center}
   \includegraphics[width=\textwidth]{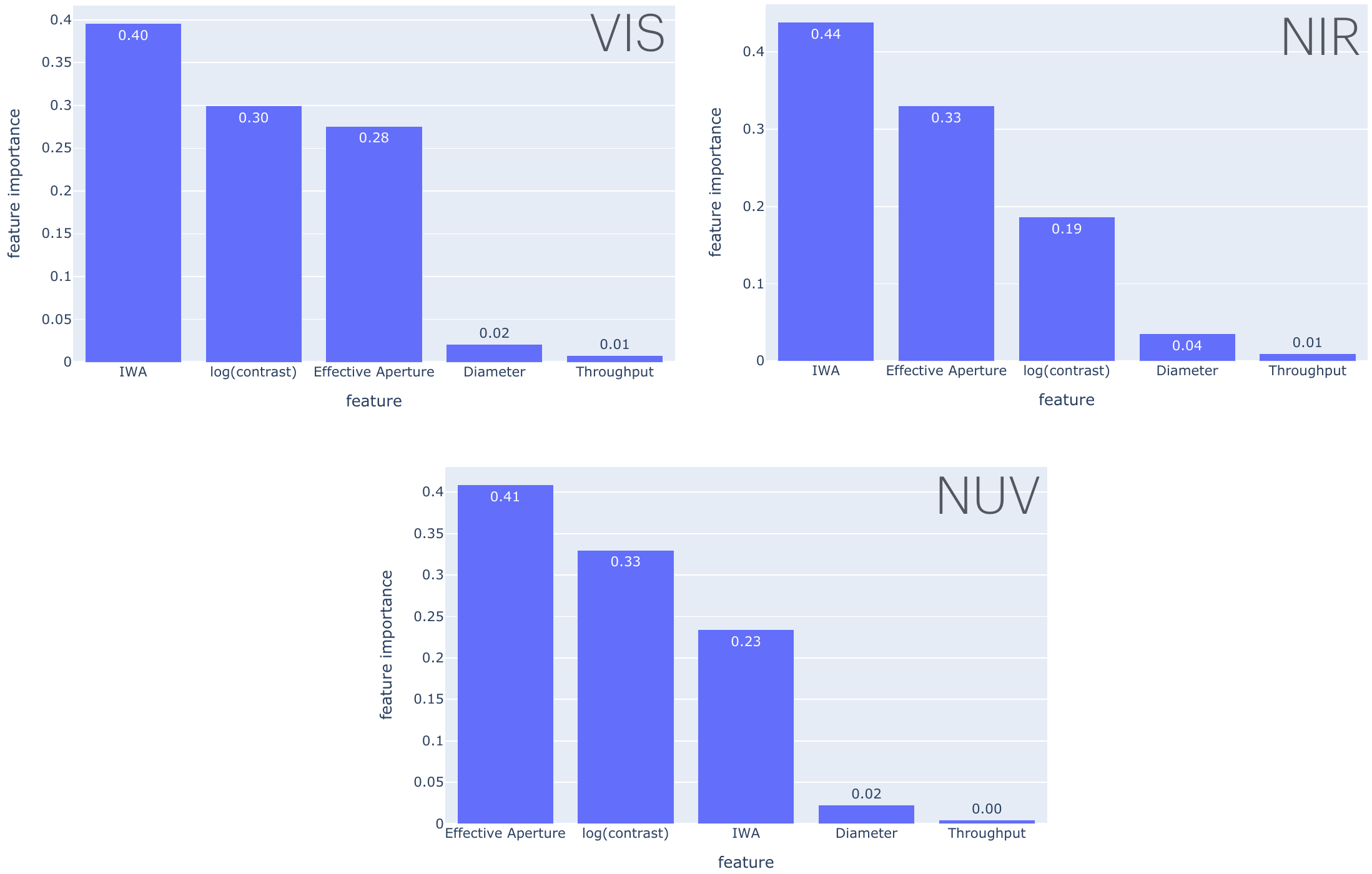}
   \end{center}
    \caption{Feature importance analysis for each of the three wavelength bands. For both VIS and NIR, the IWA is the dominant parameter that correlates with the yield. On the contrary, for NUV the effective aperture is the dominant parameter. \label{fig:feat_imp}}
\end{figure}

After a satisfactory training and evaluation of the RF\footnote{The model is available on GitHub: \href{https://github.com/MDamiano/RF_Morgan2024}{Earth-like Characterization Yield Random Forest Model}}, it is possible to explore the correlations that have been ``learned" by the RF within the dataset. The first interesting statistics to observe is the feature importance, i.e. how much each of the input parameters explains the resulting yield. Starting with the VIS band, the IWA is the dominant parameter followed by the contrast and the effective aperture at equal weight (see Figure \ref{fig:feat_imp}). Similarly, in the NIR, the IWA is dominant parameter followed by the effective aperture that appear to be more important than the contrast ratio at explaining the resulting yield. In the NUV, instead, it seems that the feature importance is different compared to the other two aforementioned bands suggesting different correlations with the yield. The effective aperture is now the dominant parameter followed by the contrast ratio and IWA lower. For all the three bands, diameter and throughput can be neglected as supported by the feature importance analysis. Physically this is also understandable as the diameter and throughput of the telescope are used in the definition of the effective aperture.

\begin{figure}[p]
   \begin{center}
   \includegraphics[width=\textwidth]{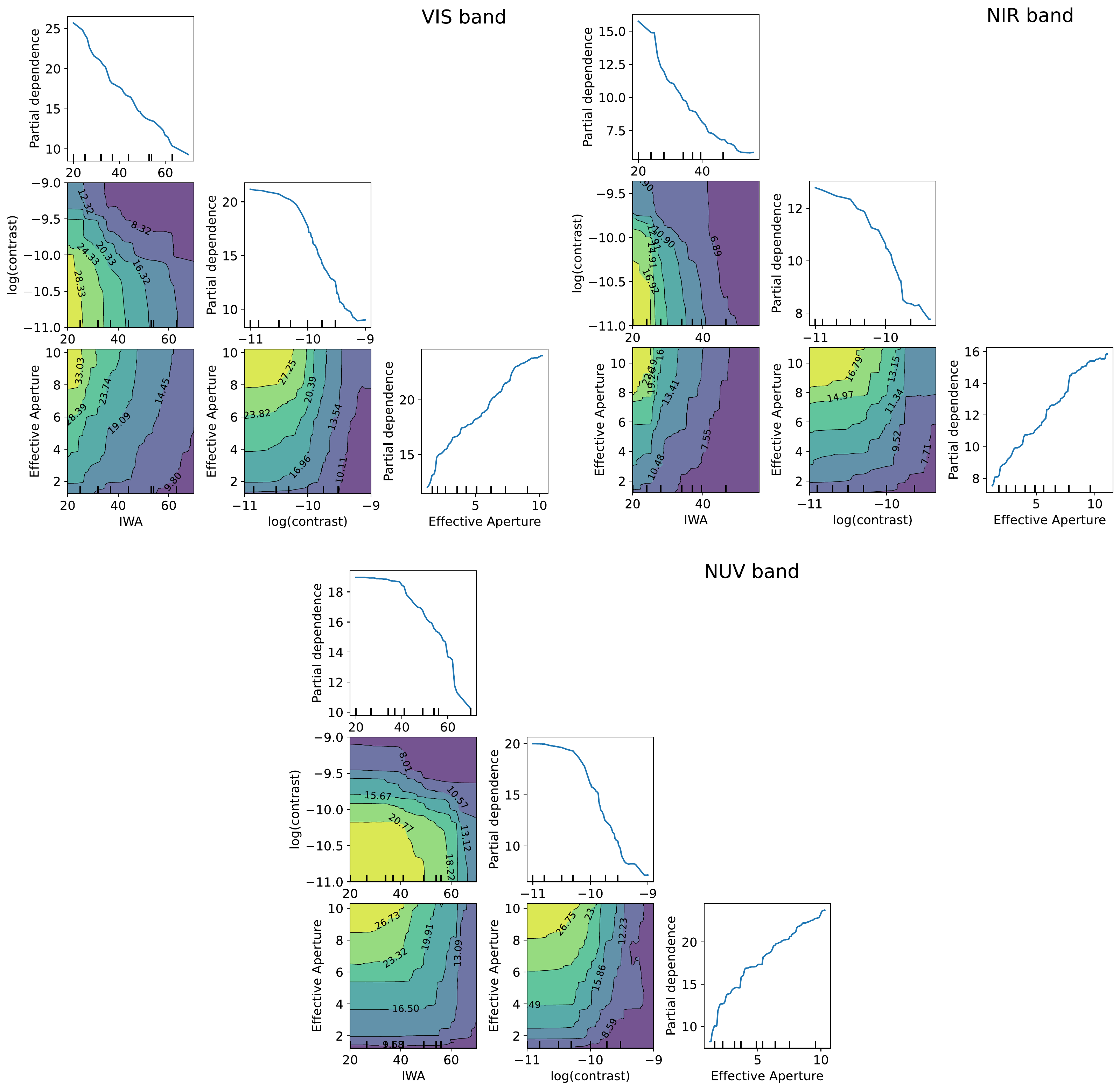}
   \end{center}
    \caption{Single and dual partial dependence plot (PDP) for each of the three wavelength bands. The diagonal of the corner plots show the single partial dependence, while the dual composed PDP is depicted in the 2D maps. The correlations between the main parameters identified by using the feature importance method are similar across the three wavelength bands. \label{fig:PDP}}
\end{figure}

Once the importance of each feature has been established, the most important parameters are then used to explore the correlations with the yield. To do so, we used the \textit{partial dependence} metric\cite{hastie2009elements}. The partial dependence reveals the dependence between the target response (i.e. resulting yield) and a set of input features of interest, marginalizing over the values of all other input features (the `complement’ features). Intuitively, we can interpret the partial dependence as the expected average resulting yield as a function of the input features of interest. 

By looking qualitatively into the three wavelength bands, the partial dependence indicates that on average the IWA and contrast have a negative linear correlation with the resulting yield (see Figure \ref{fig:PDP}). This is physically intuitive as the larger the IWA the less we can observe close to the star where Earth-like planets might orbit, and the worst the contrast ratio, the more the star will dominate the signal preventing us to see potential planets around. On the contrary the effective aperture has a positive linear correlation with\pagebreak[4] the yield on average. The larger is the effective aperture the more photons we can collect during observation campaigns.

In particular, in the VIS, by looking at the 2D correlation maps the maximum yield is reached when the effective aperture is maximized while the contrast and the IWA are minimized. Noticeably, if the contrast ratio is above 10$^{-10}$, we observe a substantial drop in yield. In the NUV, if the IWA is between 20 and 40mas and the contrast is between 10$^{-11}$ and 10$^{-10}$, the resulting yield appears to be almost insensitive. This might explain the different composition of the feature importance plot previously highlighted. Finally, the NIR, is heavily sensitive to the IWA as the partial dependence drops steadily between 20 and 40mas.


\section{CONCLUSION}
\label{sec:conclusion}
We developed a new capability to generate and run a large number of parameter sweeps for exoplanet science yield simulations. We used EXOSIMS to perform end-to-end Monte Carlo mission simulations and Dakota as a meta-wrapper around EXOSIMS to design the parameter sweep experiment and to manage execution of runs  of the 960 mission simulations. For this initial sensitivity study, we evaluated a multivariate grid in aperture diameter, coronagraph contrast, core throughput, and inner working angle. We evaluated the parameter sweep for spectral characterization in the visible, NIR, and NUV for single 20\% bandpasses. We performed a supplemental set of parameter evaluations based on the Sobel sequence that more uniformly covers the 4-dimensional parameter space than the  gridded points. The sensitivity analysis of the resultant yields used two approaches: response
surface models using polynomial fits to yield, and a nonparametric yield model using random forests (available on GitHub).

Both approaches to sensitivity analysis show that parameter interactions are important in determining\pagebreak[4] spectral characterization yield. The interaction of IWA and contrast is the most important, though the shape and strength of interaction is different for each of the three wavelength passbands. The NIR is challenging due to the IWA limitations at longer wavelengths. The NUV can access more targets due to the short wavelength of IWA (which can't be discovered by the blind-search survey in the visible) and is much more sensitive to effective aperture than Visible or NIR due to declining stellar fluxes at shorter wavelengths. For all wavelength passbands, 
contrast = $10^{-9}$ performs poorly and cannot be sufficiently compensated with other parameters; the only design point at which contrast = $10^{-9}$ can yield 25 exo-Earths is with a 9-m diameter aperture and 20 mas IWA, which is at about $1\, \lambda /D$.

At a local space closest to a nominal HWO design, yields have greater sensitivity to diameter at favorable contrast $(10^{-11})$ than at worse contrast $(10^{-9})$. Additionally, yield have a greater sensitivity to throughput at low IWA. The interaction between IWA and contrast is strong, being more sensitive to contrast in the visible and NUV but in the NIR being more sensitive to IWA.

This initial study using the our new capability for a large number of parameter sweeps focuses on the most critical parameters for a large telescope to directly image Earth-size exoplanets.  Additional parameters that come quickest to mind to explore next are contrast stability, exozodiacal light and exozodiacal light calibration error. Coronagraph performance parameters could be evaluated with real coronagraph designs or with functions quantized in working angle or with a pupil modal distribution.


\appendix    

\section{YIELD MODELING}
\label{sec:appendix}

\subsection{Astrophysical Inputs}
\label{sec:astrophysical}
The astrophysical assumptions and inputs are summarized here and discussed in greater detail in the ExSDET Final Report of the Common Comparison of HabEx and LUVOIR \cite{morgan2019sdetfinal}. The astrophysics input parameters are summarized in Table~\ref{tbl:astrophysparms}.

\begin{table}    
\caption{Adopted Astrophysical Parameters
\label{tbl:astrophysparms}}
\centering
\begin{tabularx}{.95\textwidth}{p{0.09\textwidth}p{0.255\textwidth}p{0.54\textwidth}}
\hline\hline
Parameter & Value & Description\\
\hline
$\eta_{\oplus}$ & SAG13\,power\,law & Fraction of sunlike stars w/exo-Earth candidate\\
$R_{p}$ & {$[R_p > 0.8/\sqrt{a},\,1.4]\,R_{\oplus}$} & Exo-Earth candidate planet radius\textsuperscript{a}\\
$a$ & {$[0.95,\,1.67]$\,AU} & Semi-major axis for solar twin \\
$e$ & {0} & Eccentricity (circular orbits) \\
$\cos{i}$ & {$[-1,1]$} & Cosine of inclination (uniform distribution)\\
$\omega$ & {$[0,2\pi]$} & Argument of pericenter (uniform distribution)\\
$M$ & {$[0,2\pi]$} & Mean anomaly (uniform distribution)\\
$\Phi$ & {Lambertian} & Phase function\\
$A_{G}$ & {0.2} & Geometric albedo of rocky planets\\
$A_{G}$ & {0.5} & Geometric albedo of gas planets\\
$z$  & Lindler model\textsuperscript{b} & Average V band surface brightness of zodiacal\\
$x$ & {22\,\magasec}  & V band surface brightness of 1 zodi of exozodiacal dust\textsuperscript{c}\\
$n$ & {LBTI nominal distribution} & Distribution of number of zodis for all stars\\
\hline
\multicolumn{3}{l}{\textsuperscript{a}\footnotesize{$R_p$ lower bound is $R_p > 0.8/\sqrt{a}$ in units of $R_{\oplus}$, where $a$ is the HZ-normalized semi-major axis in AU \cite{luvoir2019}.}}\\
\multicolumn{3}{l}{\textsuperscript{b}\footnotesize{Lindler zodiacal light model as a function of ecliptic latitude and longitude at observation time}}\\
\multicolumn{3}{l}{\textsuperscript{c} \footnotesize{For solar twin. Varied with spectral type, as zodi definition fixes optical depth.}}\\
\end{tabularx}
\end{table}

The exoplanet occurrence rates were based on the EXOPAG SAG-13 power law model of Kepler data, as modified by Dulz et al. \cite{DulzPlavchan2019} for large mass and large semi-major axis planets. 

Exo-Earth candidates were assumed to be on circular orbits and to reside within the conservative HZ, spanning 0.95 -- 1.67 AU for a solar twin \cite{kopparapu2013}. Exo-Earth candidates span radii ranging from $0.8a^{-0.5}$ to $1.4$ $R_{\oplus}$, where $a$ is the semi-major axis. This study focused on the exo-Earth candidate populations, using joint radius and semi-major axis distributions.

The stray light from binary stars in the final image plane was estimated \cite{sirbu2019mswc} and included as an astrophysical\pagebreak[4] noise source in exposure time calculations.  AYO and EXOSIMS  made no artificial cuts to the target list based on binarity. Instead, each code determined whether or not stray light noise made a target unobservable. 

Zodiacal cloud brightness was estimated as a function of wavelength and ecliptic latitude and longitude by interpolating published tables\cite{leinert19981997}.  EXOSIMS specifically schedules each observation, enabling it to compute the zodiacal brightness based on the target's ecliptic coordinates on the date of the observation.

The exozodiacal light level used in the yield simulations were taken from the recent results of the Large Binocular Telescope Interferometer (LBTI) survey of exozodiacal dust\cite{Ertal2018Report}. The nominal distribution function of the luminosity was used to draw an exozodiacal brightness for each star in each synthetic universe.\pagebreak[4]

We used the  ExoCat-1 catalog \cite{turnbull2015exocat}, as stored in the \verb+MissionStars+ table of the NASA Exoplanet Archive hosted by the NASA Exoplanet Science Institute (\url{https://exoplanetarchive.ipac.caltech.edu/docs/data.html}).  Missing photometric information for targets is optionally synthesized by interpolating over Eric Mamajek's Mean Dwarf Stellar Color and Effective Temperature Sequence \cite{mamajek2019modern}. ExoCat-1 is further modified in the case when the binary leakage model is being used with updated information from the Washington Double-Star catalog, maintained at the U.S. Naval Observatory.  

\subsection{Instrument Inputs} 
The instrument parameters are shown in Table \ref{tbl:instparms}.

\label{sec:instrument}
\begin{table*}
\caption{Instrument Parameters \label{tbl:instparms}}
\centering
\begin{tabular}{p{0.33\textwidth}p{0.3\textwidth}}
\hline\hline
Parameter & Value\\
\hline
Primary Diameter (m) & [6, 7, 8, 9]\\
Obscuration Factor & 0 \\
Integration Time Limit  & 60 days \\
\multicolumn{2}{c}{\underline{Starlight Suppression Performance}}\\
Raw contrast & [1e-9, 3e-10, 1e-10, 3e-11, 1e-11] \\
Raw contrast stability  & 0.1 of raw contrast\\ 
Imaging Post-processing factor & 0.29 \\
Spectra Post-processing factor & 0.1 \\
Core throughput  & [0.1, 0.2, 0.4, 0.6]\\
Photometric Aperture & 0.7 $\lambda/D$ \\
Inner Working Angle & [20, 37, 54, 70] mas \\
Outer Working Angle & 422 mas \\
Bandwidth  & 20\% \\
\multicolumn{2}{c}{\underline{Non-coronagraph Throughput }}\\
Imaging Channel & 0.29  \\
Visible Spectral Channel  & 0.135 \\
NIR Spectral Channel  & 0.167 \\
NUV Spectral Channel  & 0.145 \\
 
\multicolumn{2}{c}{\underline{Visible \& NUV Detectors}}\\
Quantum Efficiency visible & 0.9 \\
Quantum Efficiency NUV & 0.55 \\
Photon Counting Efficiency  & 0.75 \\
Dark Current (e/s)  & 3$\times 10^{-5}$\\
Read Noise (e/pix)  & 0  \\
Clock-Induced Charge (e/s) & 1.3$\times 10^{-5}$ \\

\multicolumn{2}{c}{\underline{NIR Detectors}}\\
Quantum Efficiency  & 0.7  \\
Dark Current (e/s)  & 1$\times 10^{-4}$\\
Read Noise (e/pix)  & 0 \\

\hline
\end{tabular}\end{table*}

\subsection{EXOSIMS}
\label{sec:appendix_EXOSIMS}
\begin{figure*}[ht]
    \begin{center}
         \includegraphics[trim=0 0 0 10mm,clip,width=0.9\textwidth]{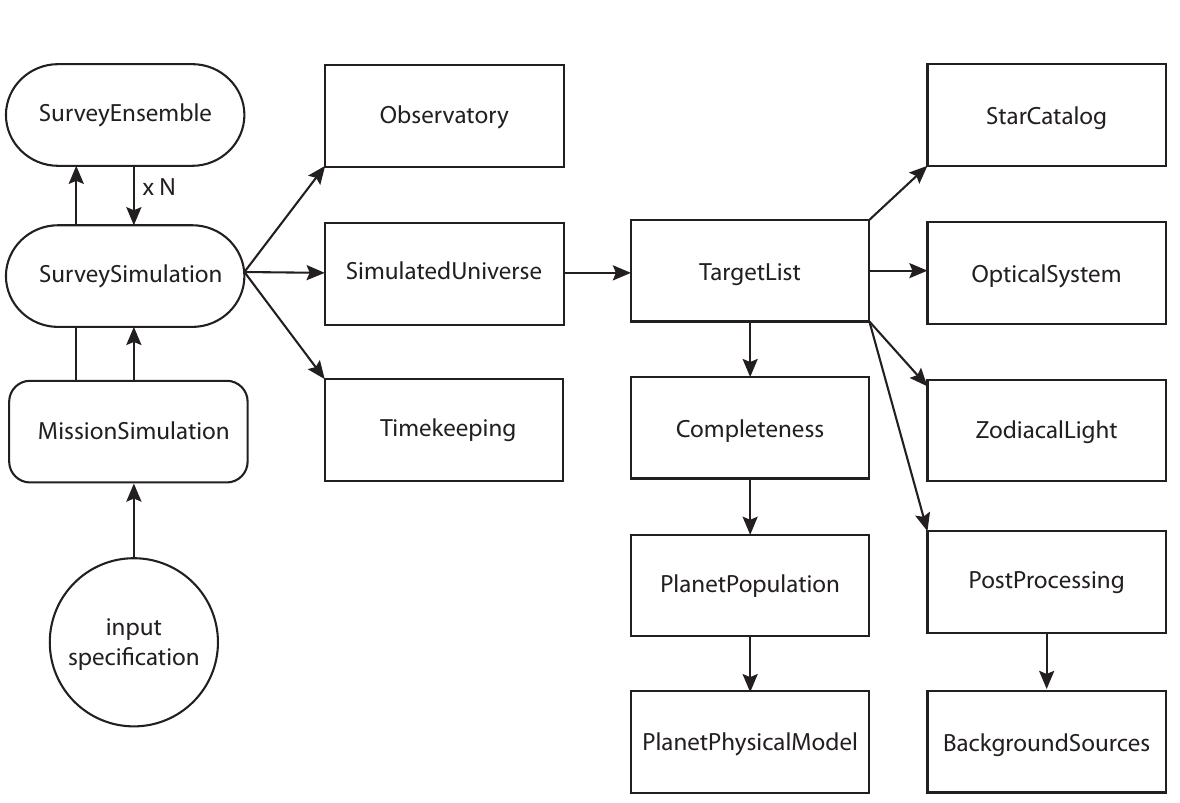}
    \end{center}
    \caption{EXOSIMS instantiation sequence, where arrows represent object instantiation. Object references are held by parent objects; for example, \textsf{TargetList} has access to both \textsf{PostProcessing} and \textsf{BackgroundSources}. Simulations described herein created a  \textsf{MissionSimulation} object, causing instantiation of \textsf{SurveySimulation}, which in turn instantiates all the other modules. 
     }
    \label{fig:instantiation_tree}
\end{figure*}

EXOSIMS is a Monte Carlo Mission Scheduling (MCMS) simulation code that simulates Monte Carlo universes. For a single universe simulation, synthetic planets are drawn from planet property distributions, and then an end-to-end mission is simulated. The mission is dynamically, sequentially scheduled and executed: the next targets to be scheduled respond to the results of previous observations (successful, null, or false positive).
   EXOSIMS consists of a collection of modules, abstracting out various state variables \pagebreak[4] and methods associated with different aspects of the full mission simulation, all with a strictly defined input/output specification.   This allows for modules to be extended, or for new modules to be implemented, without requiring any modifications to other portions of the code. EXOSIMS is thus adaptable to entirely new designs for instruments, observatories, or overall mission concepts. Figure \ref{fig:instantiation_tree} shows a schematic representation of how an EXOSIMS Mission Simulation object is constructed out of the individual modules. The module functionality split is further described in Ref.~\citeonline{morgan2021}.
   
 Yield is calculated from a generated Survey Ensemble, which contains the outputs of $N$ Survey Simulations, each of which uses exact copies of all of the downstream modules, except for the Simulated Universe.  The Simulated Universe is re-generated for each individual mission simulation by re-sampling the planet population distributions and synthesizing a new set of planets.
The execution of a single mission simulation occurs within a loop which selects 
a new target from the pool of `currently' (at the current simulated mission time) available targets, calculates the required integration time, simulates the observation and its outcome, which can be a true positive (detection), false positive (misidentified speckle or background object), true negative (null detection) or false negative (missed detection).
 
The Survey Simulation scheduler module is the backbone of EXOSIMS.  It drives the dynamic observing sequence and contains the logic for autonomous target selection.  Multiple Schedulers exist for EXOSIMS, each tailored to an architecture and an observing scenario. The Weighted Linear Cost Function (WLCF) scheduler is used for architectures having a starshade (6m C+S).  For coronagraph-only architectures, a simplified WLCF scheduler is used.

\acknowledgments 
This information is predecisional and is provided for planning and discussion purposes only. Part of this research was carried out at the Jet Propulsion
Laboratory, California Institute of Technology, under a contract with the National Aeronautics and Space Administration (80NM0018D0004). \textcopyright 2024. All rights reserved.  

\bibliography{report} 

\begin{thebibliography}{10}

\bibitem{morgan2023exo}
Morgan, R., Savransky, D., Damiano, M., Lisman, D., Mennesson, B., Mamajek, E.~E., Robinson, T.~D., and Turmon, M., ``Exo-{Earth} yield of a 6m space telescope in the near-infrared,'' in [{\em Techniques and Instrumentation for Detection of Exoplanets XI}{\nolinebreak\hspace{0.1em}]},   {\bf 12680},  518--533, SPIE (2023).

\bibitem{morgan2024exo}
Morgan, R., Savransky, D., Damiano, M., Hu, R., Mennesson, B., Mamajek, E., Turmon, M., Tokadjian, A., and Robinson, T., ``Exo-earth yield for habitable worlds observatory in the near-ultraviolet,'' in [{\em American Astronomical Society Meeting Abstracts}{\nolinebreak\hspace{0.1em}]},   {\bf 56}(2),  249--06 (2024).

\bibitem{stark2015lower}
Stark, C.~C., Roberge, A., Mandell, A., Clampin, M., Domagal-Goldman, S.~D., McElwain, M.~W., and Stapelfeldt, K.~R., ``Lower limits on aperture size for an exoearth detecting coronagraphic mission,'' {\em The Astrophysical Journal}~{\bf 808}(2),  149 (2015).

\bibitem{luvoir2019}
{The LUVOIR Team}, ``{The LUVOIR Mission Concept Study Final Report},'' {\em arXiv e-prints} ,  arXiv:1912.06219 (Dec. 2019).

\bibitem{morgan2019sdetfinal}
{Morgan}, R., {Savransky}, D., {Stark}, C., and {Nielsen}, E., ``The standard definitions and evaluation team final report: {A} common comparison of exoplanet yield.'' \url{https://exoplanets.nasa.gov/system/internal_resources/details/original/1434_Standards_Team_Final_Report_20191007.pdf} (2019).

\bibitem{damiano2021reflected}
Damiano, M. and Hu, R., ``Reflected spectroscopy of small exoplanets i: {D}etermining the atmospheric composition of sub-neptunes planets,'' {\em The Astronomical Journal}~{\bf 162}(5),  200 (2021).

\bibitem{damiano2022reflected}
Damiano, M. and Hu, R., ``Reflected spectroscopy of small exoplanets ii: {C}haracterization of terrestrial exoplanets,'' {\em The Astronomical Journal}~{\bf 163}(6),  299 (2022).

\bibitem{damiano2023reflected}
Damiano, M., Hu, R., and Mennesson, B., ``Reflected spectroscopy of small exoplanets. iii. {P}robing the {UV} band to measure biosignature gases,'' {\em The Astronomical Journal}~{\bf 166}(4),  157 (2023).

\bibitem{DKS2013qmc}
Dick, J., Kuo, F.~Y., and Sloan, I.~H., ``High-dimensional integration: The quasi-{M}onte {C}arlo way,'' {\em Acta Numerica}~{\bf 22},  133–288 (2013).

\bibitem{habexreport2020}
{Gaudi}, B.~S. et~al., ``{The Habitable Exoplanet Observatory (HabEx) Mission Concept Study Final Report},'' {\em arXiv e-prints} ,  arXiv:2001.06683 (Jan. 2020).

\bibitem{exosims-github}
Savransky, D. and the {EXOSIMS}~team, ``Exoplanet open-source imaging mission simulator.'' \url{https://github.com/dsavransky/EXOSIMS} (2020).

\bibitem{morgan2021}
Morgan, R.~M., Savransky, D., Turmon, M.~J., Mennesson, B., Dula, W., Keithly, D.~R., Mamajek, E.~E., Newman, P., Plavchan, P., Robinson, T.~D., Roudier, G., and Stark, C.~C., ``{Faster Exo-Earth yield for HabEx and LUVOIR via extreme precision radial velocity prior knowledge},'' {\em Journal of Astronomical Telescopes, Instruments, and Systems}~{\bf 7}(2),  021220 (2021).

\bibitem{Dakota}
Adams, B.~M., Bohnhoff, W.~J., Dalbey, K.~R., Ebeida, M.~S., Eddy, J.~P., Eldred, M.~S., Hooper, R.~W., Hough, P.~D., Hu, K.~T., Jakeman, J.~D., Khalil, M., Maupin, K.~A., Monschke, J.~A., Prudencio, E.~E., Ridgway, E.~M., Robbe, P., Rushdi, A.~A., Seidl, D.~T., Stephens, J.~A., Swiler, L.~P., and Winokur, J.~G., ``Dakota 6.20.0,'' tech. rep., Sandia National Laboratories (May 2024).
\newblock SAND2024-05812O.

\bibitem{tange_ole_2021}
Tange, O., ``Gnu parallel 20210822 (`kabul').'' \url{https://doi.org/10.5281/zenodo.5233953} (Aug. 2021).

\bibitem{brown2002poisson}
Brown, L.~D. and Zhao, L.~H., ``A test for the {P}oisson distribution,'' {\em Sankhya: The {I}ndian Journal of Statistics}~{\bf 64},  611--625 (2002).

\bibitem{BD2007}
Box, G.~E. and Draper, N.~R.,  [{\em Response Surfaces, Mixtures, and Ridge Analyses}{\nolinebreak\hspace{0.1em}]}, Wiley, second~ed. (2007).

\bibitem{beiman2001rf}
Breiman, L., ``Random forests,'' {\em Machine Learning}~{\bf 45}(1),  5--32 (2001).

\bibitem{hastie2009elements}
Hastie, T., Tibshirani, R., and Friedman, J.,  [{\em The Elements of Statistical Learning: Data Mining, Inference, and Prediction}{\nolinebreak\hspace{0.1em}]}, Springer series in statistics, Springer (2009).

\bibitem{DulzPlavchan2019}
{Dulz}, S., {Plavchan}, P., {Crepp}, J.~R., {Stark}, C., {Morgan}, R., {Kane}, S., {Newman}, P., {Matzko}, W., and {Mulders}, G., ``{Joint Radial Velocity and Direct Imaging Planet Yield Calculations: I. Self-consistent Planet Populations},'' {\em The Astrophysical Journal}~{\bf 893}(2) (2020).

\bibitem{kopparapu2013}
{Kopparapu}, R.~K., {Ramirez}, R., {Kasting}, J.~F., {Eymet}, V., {Robinson}, T.~D., {Mahadevan}, S., {Terrien}, R.~C., {Domagal-Goldman}, S., {Meadows}, V., and {Deshpande}, R., ``{Habitable Zones around Main-sequence Stars: New Estimates},'' {\em The Astrophysical Journal}~{\bf 765},  131 (3 2013).

\bibitem{sirbu2019mswc}
Sirbu, D., Belikov, R., Bendek, E., Henze, C., and Pluzhnik, E., ``{Demonstration of multi-star wavefront control for WFIRST, Habex, and LUVOIR},'' in [{\em Techniques and Instrumentation for Detection of Exoplanets IX}{\nolinebreak\hspace{0.1em}]},  Shaklan, S.~B., ed.,  {\bf 11117},  404 -- 417, International Society for Optics and Photonics, SPIE (2019).

\bibitem{leinert19981997}
Leinert, C., Bowyer, S., Haikala, L., Hanner, M., Hauser, M., Levasseur-Regourd, A., Mann, I., Mattila, K., Reach, W., Schlosser, W., et~al., ``The 1997 reference of diffuse night sky brightness,'' {\em Astronomy and Astrophysics Supplement Series}~{\bf 127}(1),  1--99 (1998).

\bibitem{Ertal2018Report}
Ertal2018Report, ``Report on hosts survey return.'' \url{http://nexsci.caltech.edu/missions/LBTI/report_hosts_dec2018_v1.2.pdf} (2018).

\bibitem{turnbull2015exocat}
Turnbull, M.~C., ``Exocat-1: The nearby stellar systems catalog for exoplanet imaging missions.'' arXiv preprint arXiv:1510.01731 (2015).

\bibitem{mamajek2019modern}
Mamajek, E., ``A modern mean dwarf stellar color and effective temperature sequence.'' \url{http://www.pas.rochester.edu/~emamajek/EEM_dwarf_UBVIJHK_colors_Teff.txt} (2019).

\end{thebibliography}
\bibliographystyle{spiebib} 

\end{document}